\documentclass[printer]{aa}

\usepackage[varg]{txfonts}
\usepackage[dvips]{color}
\usepackage{graphicx,subfigure}
\usepackage{amsmath}
\usepackage{epsfig,amsmath,amssymb}

\def\ga{\,\hbox{\hbox{$ > $}\kern -0.8em \lower 1.0ex\hbox{$\sim$}}\,}
\def\la{\,\hbox{\hbox{$ < $}\kern -0.8em \lower 1.0ex\hbox{$\sim$}}\,}
\def\beq{\begin{equation}}
\def\eeq{\end{equation}}

\begin{document}

%
\titlerunning{Shell instability of a collapsing core}
\authorrunning{Ntormousi \& Hennebelle}
\title{Shell instability of a collapsing dense core}
\author{Evangelia Ntormousi\inst{1}, Patrick Hennebelle\inst{1,2}}
\date{Received -- / Accepted --}

\institute{
Laboratoire AIM, 
Paris-Saclay, CEA/IRFU/SAp - CNRS - Universit\'e Paris Diderot, 91191, 
Gif-sur-Yvette Cedex, France \\
\and
LERMA (UMR CNRS 8112), Ecole Normale Sup\'erieure, 75231 Paris Cedex, France}

\abstract
{}
{Understanding the formation of binary and multiple stellar systems largely comes down to studying the 
circumstances under which a condensing core fragments (or not) during the first stages of the collapse.
However, both the probability of fragmentation and the number of fragments seem to be determined to a large degree by the initial conditions.  
In this work we explore this dependence by studying the fate of the linear perturbations of a homogeneous gas sphere, both analytically and numerically.}
{In particular, we investigate the stability of the well-known homologous solution that describes the collapse of a uniform spherical cloud.
One problem that arises in such treatments is the mathematical singularity in the perturbation equations, which corresponds to
the location of the sonic point of the flow.  
This difficulty is surpassed here by explicitly introducing a weak shock next to the sonic point as a natural way of connecting the subsonic to the supersonic regimes.
In parallel, we perform adaptive mesh refinement (AMR) numerical simulations of the linear stages of the collapse and compared the growth rates obtained by each method.}
{With this combination of analytical and numerical tools, we explore the behavior of both spherically symmetric
and non-axisymmetric perturbations.  
The numerical experiments provide the linear growth rates as a function of the core's initial virial parameter and as a function of the azimuthal wave number of the perturbation.  The overlapping regime of the numerical experiments and the analytical predictions is the situation of a cold and large cloud, and in this regime the analytically calculated growth rates agree very well with the ones obtained from the simulations.}
{The use of a weak shock as part of the perturbation 
allows us to find a physically acceptable solution to the equations for a continuous range of growth rates.  
The numerical simulations agree very well with the analytical prediction for the most unstable cores, 
while they impose a limit of a virial parameter of 0.1 for core fragmentation in the absence of rotation.}

\keywords{ Stars: formation - Interstellar medium:structure }

\maketitle
\section{Introduction}

By now it is well understood that stars form out of the gravitational condensation of cold and dense molecular cores, 
which are very structured in density and in velocity
\citep{Larson_1981,Falgarone_Phillips_1990}.
In parallel, it has also been argued that most stars live in binaries or multiple systems 
\citep[e.g,][]{Duquennoy_Mayor_1991,Goodwin_2007}. 
It is therefore natural to assume that owing to these internal velocities and density enhancements, 
a core typically fragments at a certain stage of its collapse
and eventually produces two or more stars.  

From an observational perspective, the matter can be elucidated by registering the degree of multiplicity
at different stages of the core contraction.  
Most data for early stellar multiplicity are available for Class I/II objects, in other words, for protostars with almost no 
trace of their initial envelope left \citep{Ghez_93,Patience_2002,Duchene_2004,Duchene_2007}.
But, to clarify the dependence on the initial physical
properties and internal structure of a core, one must allude to even earlier phases in the collapse
(Class 0 objects),
when the protostars are still largely embedded in their parental core \citep{Andre_93,Andre_00}.
Although challenging because of the obscuration from the envelope, 
there have been numerous attempts to constrain stellar multiplicity that early in the collapse stage of a core
 \citep{Chen_08,Girart_09,Duchene_2004,Duchene_2007,Jorgensen_07}.
The results show an increase in the number of fragments from the Class 0 to the Class I stage, 
which is evidence that suggests that fragmentation is favored either in the Class 0 stage or very shortly afterward \citep{Maury_10}.

There have been also numerous theoretical studies, both of the collapse process
and of the fragmentation during collapse.
\citet{Ebert_1955} and \citet{Bonnor_1956} independently derived an analytic criterion for a static isothermal sphere to be stable against gravitational collapse. In their numerical simulations of collapsing, initially uniform isothermal spheres \citet{Larson_1969} and \citet{Penston_1969} found a one-dimensional, self-similar solution (called LP solution or flow in the following).  If x is the self-similar variable of the profile, this solution describes a cloud with a flat density profile that turns into a $x^{-2}$ dependence toward infinity and a velocity profile proportional to x.  It was later discovered by \citet{Shu_1977} that this solution is a member of a whole family of self-similar collapse solutions, many of which contain critical points.
In that paper, he arrived at the conclusion that an initially centrally condensed cloud will collapse from the inside out, establishing an expansion wave.
This behavior was integrated into a more complete picture of the collapse and the behavior of self-similar collapse solutions by \cite{Whitworth_Summers_1985}.

The stability of the collapse solutions is understood to a much lesser extent. 
For the case of a static, uniform, presureless spherical cloud, \citet{Hunter_1964} showed
that there is an unstable shell mode that grows like $(t_0-t)^{-1}$, where $t$ is the time and $t_0$ the time at which collapse occurs. 
 
\citet{Hanawa_Matsumoto_1999} (HM99 in the following) performed a stability analysis of the LP flow using a shooting method.
In this type of analysis, one starts integrating the equations from one boundary and varies some parameters until the solution at the other boundary matches the desired conditions there.
They came to the conclusion that the solution is stable overall, with the exception of 
a slowly growing $l=2$ mode (where l the azimuthal wave number of a spherical harmonic perturbation) with a growth rate
of 0.354.  
This very weak instability is consistent with the analysis by \citet{Ori_Piran_1988}, who derived a stability criterion for
self-similar isothermal collapse flows that is based on the gradient of the radial velocity close to the critical point.
According to that criterion, a homogeneous isothermal sphere is an unstable configuration that could naturally
converge to a (generally much stabler) Larson-Penston type flow.  

While quite interesting, this result poses a few questions. 
One is that the growth rate is rather slow, so it is unclear whether perturbations can really 
grow sufficiently (see Sect.~\ref{discussion}).  Another issue is that
only the $l=2$ mode was found to be unstable, since HM99 report the nonexistence of eigenvectors for higher values of $l$. 
Finally, no total eigenvector has been calculated for the spherical mode.  Altogether, this suggests 
that the stability analysis of a collapsing cloud is not yet complete.

On the other hand, there is abundant literature on numerical simulations of the fragmentation of rotating and/or turbulent cores.
\citet{Boss_Bodenheimer_1979} studied the fate of an $l=2$ perturbation on a collapsing core, and \citet{Bodenheimer_Burkert_1993}
repeated the experiment, finding that a filament connecting the fragments should develop .
(This experiment has been repeated many times since and is used extensively as a code testing tool.)
A criterion for fragmentation in the presence of rotation was provided by the semi-analytical work of \citet{Tohline_1981}, 
quantifying thermal support with the virial parameter $\alpha$ and the the rotational versus gravitational energy in the cloud with the 
corresponding parameter $\beta$.  The quantity $\alpha\beta$ has since been used repeatedly to delimit the conditions for fragmentation \cite[for instance]{Miyama_1984, Hachisu_Eriguchi_1984, Tsuribe_Inutsuka_1999, Tohline_2002}, and it was found that $\alpha\beta<0.1-0.2$ typically leads to fragmentation. 

At the same time, increasingly more complex direct simulations
of the collapse and fragmentation processes have been performed, 
including the internal and external thermal pressure of
the core, rotation \citep{Myhill_Kaula_1992,Cha_Whitworth_2003,Matsumoto_Hanawa_03,Hennebelle_2004, Machida_2008, Commercon_2008}, 
magnetic fields \citep{Banerjee_Pudritz_2006,Price_Bate_2007,Hennebelle_Teyssier_08,Boss_Keiser_2013}, and turbulence 
\citep{Klessen_1998, Klessen_01, Offner_08, Bate_09, Joos_13}, which seem to affect the number of fragments and their separation.  
\citet{Girichidis_11} conducted a parameter
study in which they varied the initial density profile of the core and the level of turbulence. They provide evidence of the strong
 influence of the mean radial density profile to the result of the collapse.

Given the abundance and complexity of the existing models for fragmentation, it is somewhat
surprising that there is so little to be said for the stability of a solution that is as simple as the homologous, uniform, isothermal collapse.
It is clear that this solution has limited applicability to real cores, which are sometimes far from isothermal spheres and do host turbulent motions and magnetic fields, which are all potentially important and complex effects that can  substantially alter the behavior of the solution. 
However, a study of its linear stability offers important insight into the principal mechanisms causing fragmentation, since it does not suffer from the complexity of nonlinear effects and can contribute to the overall understanding of core collapse.

In the present paper we rectify by performing a linear stability analysis of this flow. 
Our study shows that the homologous solution is indeed unstable and we obtain the corresponding growth rates.  
This analysis is presented in Sections~\ref{sec:stability} and ~\ref{sec:nonaxissymm}.  
In the fourth section we present numerical simulations of a uniform 
dense core subject to a spherical perturbation and we look for the range of parameters for which it is unstable.  
We measure the growth rate of the shell mode (a spherically symmetric perturbation), 
as well as higher-order spherical harmonic perturbations and find them to be in good agreement 
with the result of our linear analysis. The fifth section discusses 
the implications of our results and proposes further interpretation.  The sixth section concludes the paper. 

\section{Stability analysis: the spherical case}
\label{sec:stability}

\subsection{Equations and Self-similarity}

We investigate the stability of a collapsing isothermal sphere against linear perturbations.
Given the symmetry of the problem, we start with the equations of hydrodynamics in spherical coordinates
in one dimension:

\begin{flalign}
   \partial_t\rho + \frac{\partial_r(r^2\rho u_r)}{r^2}  =0,  \nonumber \\
   \partial_t u_r + u_r\partial_r u_r  =  
 - C_s^2 \frac{\partial_r\rho}{\rho} - \partial_r\psi, \label{eq_hydro} \\
   \frac{1}{r^2} \partial_r (r^2\partial_r \psi)  = 4 \pi G  \rho,
\nonumber
\end{flalign}
where $\rho$ the gas mass density, $u_r$ the radial velocity, and $\psi$ the gravitational potential.
It is well known \citep{Larson_1969,Penston_1969,Shu_1977} that 
this system admits self-similar solutions of the form
\begin{eqnarray}
   X = {r \over C_s (t_0-t)},  \nonumber \\
   {\cal{R}}(X) = 4 \pi G (t_0-t)^2 \rho(r,t),  \label{self_sim} \\  
   {\cal{U}}(X) = {u_r(r,t) \over C_s}, \nonumber \\
   \Phi (X)  = { \psi(r,t) \over C_s^2},  \nonumber
\end{eqnarray}
where the density ${\cal{R}}$, the radial velocity ${\cal{U}}$, and the gravitational potential, $\Phi$ are functions of the self-similar variable $X$, $C_s$ is the sound speed, and G the gravitational acceleration.
Among the various solutions known in the literature the simplest is the homologous solution
\begin{eqnarray}
   R_0 = {2 \over 3},  \nonumber \\
   U_0= \alpha_0 X  = -{2 \over 3} X, \label{homol} \\
   \phi_0 = {R_0 \over 6} X^2, \nonumber
\end{eqnarray}
which describes a cloud with a uniform density $R_0$ and a velocity field proportional to $X$ with a slope $\alpha_0$.
Clearly, like other self-similar profiles, this solution
is only valid within certain spatial and temporal limits, as both the mass and the velocity diverge for large values of $X$.
In addition, in a real collapse situation there is a rarefaction wave propagating outwards \citep{Tsuribe_Inutsuka_1999,Truelove_1997},
 which breaks the self-similarity.  In order then for this solution to remain valid the gas has to be cold enough for the collapse to happen before the rarefaction wave can reach the center.  Then Eqs. (\ref{homol}) can be used to describe the inner parts of the cloud.

The simplicity of this profile and the fact that it describes the center of a cold core quite accurately make it a popular initial condition for numerical simulations \citep[for example]{Larson_1969,Price_Bate_2007,Commercon_2008,Boss_Keiser_2013}. It is therefore both interesting and relevant to perform a stability analysis of this solution and to confront the analytical results with numerical simulations.

We first look for perturbations of the form 
\begin{eqnarray}
   R(r,t) = R_0(X) + \delta R(X) (t_0-t)^{-\sigma} , \nonumber \\ 
   U(r,t) = U_0(X) + \delta U(X) (t_0-t)^{-\sigma}, \label{perturb} \\
   \phi(r,t) =  \phi_0(X) +  \delta \phi (X) (t_0-t)^{-\sigma}, \nonumber 
\end{eqnarray}
where $R_0$, $U_0$, and $\phi_0$ denote the equilibrium state and $\delta R(X)$, $\delta U(X)$, and $\delta\phi (X)$ the perturbed quantities.
The quantity $\sigma$ in the exponent is assumed to be positive and represents the growth rate of the perturbation.
Inserting these expressions into Eqs.~(\ref{eq_hydro}), we get 
\begin{small}
\begin{eqnarray}
   (\sigma +2) \delta R + X \delta R ' + {R_0 \over X^2} \partial_X (X^2  \delta U) + {\alpha_0 \over X^2 } \partial_X(X^3 \delta R) = 0, \label{eq_pert1} \\
   (\sigma + \alpha_0) \delta U + (1 + \alpha_0) X  \delta U ' = - { \delta R' \over R_0} -  \delta \phi '  , \label{eq_pert2} \\
   \delta \phi '' + {2 \delta \phi' \over X } = \delta R. \label{eq_pert3} 
\end{eqnarray}
\end{small}
with primes denoting the spatial derivatives.
Combining Eqs.~(\ref{eq_pert1}) and~(\ref{eq_pert3}), one can show that 
\begin{eqnarray}
   \delta\phi ' =   {1 \over 1 - \sigma } \left( R_0 \delta U + (1+\alpha_0) X \delta R  \right) + {K \over X^2},
   \label{eq_psi}
\end{eqnarray}
where $K$ is a constant chosen to ensure the continuity of $\delta\phi$. 
From Eqs.~(\ref{homol}),~(\ref{eq_pert1}),~(\ref{eq_pert2}),  we get 
\begin{eqnarray}
   \delta R' =  { 1 \over 3 - X^2/3} \left( \sigma X \delta R + (8/3 - 2 \sigma) \delta V - 2 \delta \phi'   \right) ,  
   \label{eq_dR}  
\end{eqnarray}
which together with Eqs.~(\ref{eq_pert2}) and~(\ref{eq_psi}) allows us to solve the system. 
Once the boundary conditions have been specified, a numerical integration is carried out by a standard Runge-Kutta scheme. 

\subsection{Inner boundary and the critical point}

The boundary conditions for the perturbation at $X \simeq 0$ are given by
\begin{eqnarray}
   \nonumber
   \delta R &\rightarrow& R_1 ={\rm cst} , \\
   \delta U &\rightarrow& -{\sigma R_1 \over 2} X.
   \label{bound_cond}
\end{eqnarray}
where cst is a constant. So by simply specifying a constant value for $R_1$ we can integrate the system of Eqs.~(\ref{eq_pert2}) and~(\ref{eq_psi}) for different values of $\sigma$.

One complication to this otherwise straightforward approach 
is that the system possesses a critical point at $X_{\rm crit} = 3$, where the gas becomes supersonic with respect to the self-similar profile. 
The presence of such a boundary poses a restriction to our numerical solutions, since only the ones which cross it can be considered physically acceptable.

It is easy to show that the system ~(\ref{eq_pert1})-(\ref{eq_pert3}) has two exact solutions that 
resemble these boundary conditions and are given by 
\begin{eqnarray}
   \delta R = R_1 = {\rm cst} , \\
   \delta U = -{\sigma R_1 \over 2} X, \\
   \sigma = 1 \; {\rm or} \; -{2 \over 3}. 
   \label{exact_sol}
\end{eqnarray}
While mathematically acceptable, these solutions offer little physical
insight: they merely describe a variation of the mean density.
They are, however, mathematically acceptable, since they remain small at any $X$ with respect to the perturbed solutions. 

Apart from these trivial forms, the integrations of the system of Eqs.~(\ref{eq_pert2}) and~(\ref{eq_psi}) we performed for different values of $\sigma$ gave no other solutions able to cross the critical point. 

\subsection{Shock conditions}
\label{shock}

From the above discussion it becomes clear that a more general form of the perturbation should exist which allows
for physically meaningful solutions to the corresponding perturbation equations and which does not suffer from the 
issue of critical point crossing.

To begin with, one can imagine a core whose size is much larger than the sonic radius.  
In this case the perturbation can be restricted to the supersonic parts, where $X$ is much greater than the critical value.  
For a cloud this large we are essentially in the same regime as the cold cloud described by the homologous collapse solution (Eqs. (\ref{homol})), where gravitational collapse happens before thermal pressure effects have had time to break the self-similarity.  We expect then that the behavior of the perturbations will be similar to the general $t^{-1}$ growth found by \citet{Hunter_1964} for the case of a cold cloud.

In order to integrate the solution for the outer parts of the cloud, we need a physically meaningful boundary condition at the vicinity of the critical point.  We thus seek solutions that include a shock. In the vicinity
of the critical point, the velocity of the fluid is by definition transonic with 
respect to the self-similar profile, so an arbitrary weak shock can naturally happen in this area.  
Note that, although the spatial derivatives of the density and velocity fields are infinite at the 
shock, all fields remain small if the shock is weak.  Therefore, such a discontinuity indeed generates a linear perturbation. 

Let the shock be located at $X_{\rm{shock}} = X_{\rm crit} + \delta X$ where $\delta X >0$.
This is the location where the unperturbed solution, given by 
$\delta R=0$, $\delta U=0$ is connected to a state given by the Rankine-Hugoniot (RH) conditions, which in the frame of reference of the shock are
\begin{eqnarray}
   \left( u_{r,1}-V_{\rm shock} \right) \left( u_{r,2}-V_{\rm shock} \right) = C_s^2, \nonumber \\
   {\rho_2 \over \rho_1} = {(u_{r,1}-V_{\rm shock})^2 \over C_s^2}. 
   \label{rankine}
\end{eqnarray} 
where $V_{\rm shock}$ is the velocity of the shock and subscripts 1 and 2 mark the pre- and post- shock quantities, respectively. 

The velocity of the flow with respect to the self-similar profile is 
given by $\alpha_0 X + X= X/3$.  Thus at $X_{\rm{shock}}$, 
the Mach number can be expressed as ${\cal{M}}=X_{\rm shock}/3 = 1 + \delta X /3$.
In the frame of the shock, the gas from the subsonic region is entering
supersonically into the shock and therefore constitutes the pre-shock medium whose density is amplified.

Since we consider that the subsonic region is unperturbed, 
the expressions for $\rho_1$ and $u_{r,1}$ that enter the jump conditions in Eqs.~(\ref{rankine}) are simply given by Eqs.~(\ref{self_sim}) and~(\ref{homol}). 

By replacing we get 
\begin{eqnarray}
   \left( 1 + {\delta X  \over 3} \right) \left( 1 + {\delta X  \over 3} + \delta U (t_0-t)^{-\sigma} \right) = 1, \nonumber \\
   {\delta R (t_0-t)^{-\sigma} +  R_0 \over R_0 } = \left( 1 + {\delta X  \over 3} \right)^2 . 
   \label{rankine2}
\end{eqnarray} 
At the limit of a weak shock, we can expand these relations and get 
\begin{eqnarray}
  \delta U (t_0-t)^{-\sigma} = -  {2   \over 3} \delta X, \nonumber \\
  \delta R (t_0-t)^{-\sigma}  =  {2 R_0 \delta X  \over 3} =  {4    \over 9} \delta X. 
\label{rankine_expan}
\end{eqnarray} 
As the perturbation develops it moves toward larger $\delta X$, so the density increases and the shock becomes stronger. 

Since the perturbed density remains zero up to the position of the shock, 
the constant $K$ in Eq.(\ref{eq_psi}) must be chosen in such a way that $\delta \phi' (X_{\rm shock})=0$.

\subsection{Limit at large radii}

In order to identify the physically relevant solutions, the asymptotic behavior at large $X$ must also been known. 
From Eqs.~(\ref{eq_pert1})-(\ref{eq_pert3}), it is easy to infer that there must be an asymptotic 
behavior of the form
\begin{eqnarray}
\delta R &\rightarrow& R_\infty X^{-v-1}, \label{asym1} \\
\delta U &\rightarrow& U_\infty X^{-v}. 
\label{asym2}
\end{eqnarray}
where the exponent $v$ in Eqs.~(\ref{asym1}) and (\ref{asym2}) should be chosen such that the asymptotic form satisfies Eqs.~(\ref{eq_pert1})-(\ref{eq_pert3}). 
By inserting expressions (\ref{asym1}) and (\ref{asym2}) into Eqs.~(\ref{eq_pert1})-(\ref{eq_pert3})
and dropping the thermal pressure term, which is negligible at large $X$, we obtain two linear 
equations for $R_\infty$ and $U_\infty$, which depend on 
$\sigma$ and $v$. By demanding that the determinant of this 
system be zero, we get the two solutions for $v$:
\begin{eqnarray}
   v &=& 3 \sigma - 4, \nonumber \\
   v &=& 3 \sigma + 1.
   \label{deter}
\end{eqnarray}
We thus recover the exact homologous solution mentioned above,
since when $\sigma=1$ or $-2/3$, we get $v=-1$, which means $U \propto X$
and $R={\rm cst}$. 
On closer inspection we also see that the first branch ($v = 3 \sigma - 4$) diverges most rapidly.
Since mathematically it is required that $v > -1$ for the solutions to behave well towards infinity, it follows that 
$\sigma > 1$. When $1 < \sigma < 4/3$, the velocity diverges at
infinity but still remains everywhere negligible with respect to the 
perturbed solution. 

\subsection{Numerical integration}

The shock is now essentially the low-$X$ boundary for the numerical integration, 
so we only need to specify its position $\delta X$ from the critical point in order to define the boundary conditions.  
This value must be small in order to ensure linearity. 

As shown by Eqs.~(\ref{rankine_expan}), the exact values of  $\delta U$
and $\delta R$ are not important (they depend on $t_0$, which can be freely chosen)
but their ratio is fixed.  

Starting from a location close to zero (specifically $X_{s}=10^{-2}$), and using a step $d X=10^{-3}$, we integrate the perturbation equations up to $X=100$.
We have checked that the solutions are well resolved and converged.

\setlength{\unitlength}{1cm}
  \begin{figure}
    \begin{picture} (0,14)
    \put(0,7){\includegraphics[width=9cm]{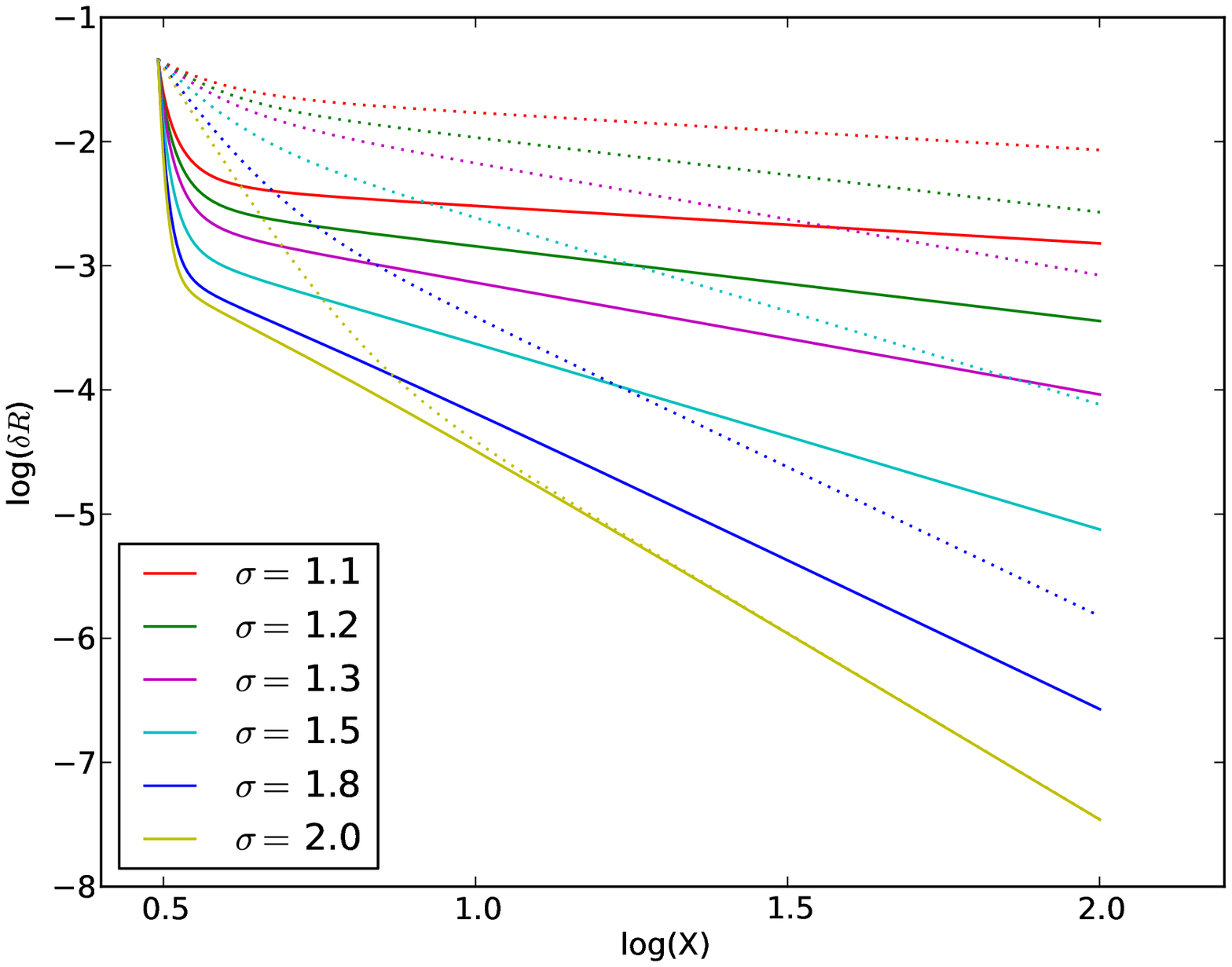}}
    \put(0,0){\includegraphics[width=9cm]{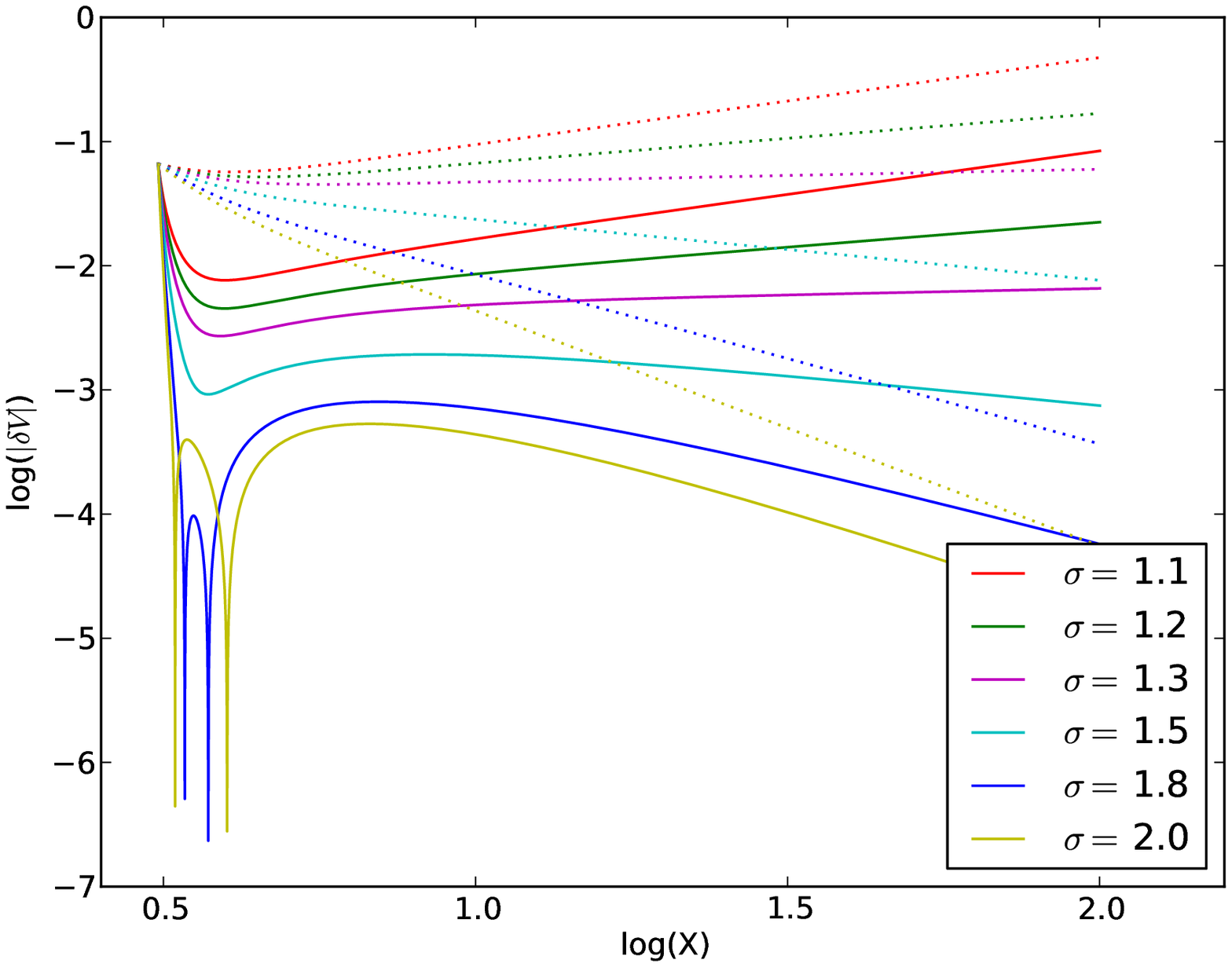}}
  \end{picture}
  \caption{Density (top panel) and velocity (bottom panel) fields of the solutions for various values of $\sigma$.
For values of $X$ smaller than the position of the shock ($X<X_{sh} \simeq 3$), the density and the velocity are unperturbed, 
i.e. $\delta R=\delta V=0$.  The solid lines show the solutions for the system of Eqs.~(\ref{eq_pert1})-(\ref{eq_pert3})
For comparison, solutions to the same equations but without the thermal pressure term are shown with the dotted lines.}
  \label{res_dens}
\end{figure}

Figure~\ref{res_dens}  shows the density and the velocity fields for various
values of the growth rate, $\sigma$. The solid lines show the solutions of 
Eqs.~(\ref{eq_pert1})-(\ref{eq_pert3}) and the dotted lines show the solutions
for the same system of equations when the thermal pressure is suppressed. 
The latter case  corresponds to a cold cloud and is given for 
comparison. 

The velocity and the density fields vary rapidly in the vicinity of the shock, 
$X \simeq 3$. The velocity, which is negative, increases and the density decreases.
At $X \simeq 4-5$, the behavior of the density changes: it starts decreasing much less 
rapidly and it quickly reaches the asymptotic behavior expressed by 
Eqs.~(\ref{asym1} and \ref{asym2}). 

The behavior of the velocity is slightly more 
complex. For $\sigma < 4/3$, it decreases for $X > 4-5$ as expected from
the asymptotic form (Eqs.~\ref{asym1} and \ref{asym2}). 
For $4/3 < \sigma < 1.6$, the velocity eventually decreases and tends towards zero after 
a small increase. Again, this is in good agreement with the expected asymptotic profile. 
In contrast, for larger values of $\sigma$, the behavior is identical ,only the velocity becomes positive just after the shock 
at $X \simeq 3-5$, which means that the perturbation is expanding. 

Based on this form of the perturbations, we believe that the regime $1 < \sigma < 4/3$ 
corresponds to the most physically relevant solutions. For smaller $\sigma$, the velocity diverges 
at large $X$ and is therefore not really a perturbation with respect to the unperturbed solution.
When $\sigma > 4/3$, the velocity quickly tends to zero, meaning that it is very localized around the shock. 
Finally, when $\sigma > \simeq 1.7$, the velocity becomes positive.  This situation
probably corresponds to somehow artificial perturbations that are unlikely to represent
a physically relevant case. 

This conclusion remains almost identical for the cold case since for 
$\sigma < 4/3$ the velocity diverges toward large $X$ while for $\sigma > \simeq 1.8$, 
the density becomes negative at $X > \simeq 5$. \\ 

To summarize, we find that linear perturbations that include a weak shock in the 
vicinity of the sonic point present physically acceptable behavior if the 
 growth rate is larger than unity.  As the growth rate becomes larger than $4/3$, the form of the eigenvectors
suggests that the corresponding perturbation is unlikely to occur: The velocity field 
rapidly converges toward zero, implying that the perturbation is isolated in the sonic 
region. 

\section{Stability analysis: the non-axisymmetric modes}
\label{sec:nonaxissymm}

The instability of a shell-like structure is an important result, but if we care about
the formation of multiple fragments we must also consider non-spherically shaped perturbations.
In this section we present the stability analysis for non-spherically symmetric modes.
Since the method is practically the same as the spherical case, here we will only highlight the differences. 

\subsection{Equations}
The fluid equations for a self-gravitating gas in spherical coordinates and three dimensions are:

\begin{flalign}
   \partial_t\rho + \frac{\partial_r(r^2\rho u_r)}{r^2}  + {1 \over r \sin \theta } \partial _\theta ( \sin \theta \rho u_\theta)
 + {1 \over r \sin \theta } \partial _\phi ( \rho u_\phi)  =0,  \nonumber \\
   \partial_t u_r + u_r\partial_r u_r + { u_\theta \over r} \partial_ \theta u_r + {u_\phi \over r \sin \theta} \partial_ \phi  u_r 
 - { u_\theta^2 + u_ \phi^2 \over r} =  \nonumber \\
 - C_s^2 \frac{\partial_r\rho}{\rho} - \partial_r\psi,  \label{eq_hydro_ns} \\
   \partial_t u_\theta + u_r\partial_r u_\theta + { u_\theta \over r} \partial_ \theta u_\theta + {u_\phi \over r \sin \theta} \partial_ \phi  u_\theta 
   + { u_r u_\theta   \over r}  - { u_ \phi^2 \over r \tan \theta} = \nonumber \\  
 - {C_s^2 \over r } \frac{\partial_ \theta \rho}{\rho} - {1 \over r} \partial_\theta \psi,  \nonumber \\
   \partial_t u_\phi + u_r\partial_r u_\phi + { u_\theta \over r} \partial_ \theta u_\phi + {u_\phi \over r \sin \theta } \partial_ \phi  u_\phi 
   + { u_r u_\phi   \over r}  = \nonumber \\  
 - {C_s^2 \over r \sin \theta} \frac{\partial_ \phi \rho}{\rho} - {1 \over r \sin \theta} \partial_\phi \psi,  \nonumber \\
   \frac{1}{r^2} \partial_r (r^2\partial_r \psi)  + \frac{1}{r^2 \sin \theta} \partial_\theta ( \sin \theta \partial_\theta \psi) 
+ \frac{1}{r^2 \sin^2 \theta}  \partial^2 _{\phi^2} \psi = 4 \pi G  \rho, \nonumber
\end{flalign}
%
where $u_r$, $u_{\theta}$, and $u_\phi$ denote the velocities along each of the spherical coordinates $r$, $\theta$ and $\phi$.
To study the stability of the homologous solution with respect to non-spherical modes,
 we look for perturbations of the form (HM99):

\begin{eqnarray}
   R(r,\theta,\phi,t) = R_0(X) + \delta R(X) Y_l^m (\theta,\phi) (t_0-t)^{-\sigma} , \nonumber \\ 
   U(r,\theta,\phi,t) = U_0(X)+ \delta U(X) Y_l^m (\theta,\phi) (t_0-t)^{-\sigma}, \nonumber \\ 
   V(r,\theta,\phi,t) = \delta V(X) {1 \over l+1} \partial_\theta Y_l^m (\theta,\phi) (t_0-t)^{-\sigma}, \label{perturb_ns}  \\  
   W(r,\theta,\phi,t) = \delta V(X) {1 \over l+1} {1 \over \sin \theta} \partial_\phi Y_l^m (\theta,\phi) (t_0-t)^{-\sigma}, \nonumber \\ 
   \phi(r,\theta,\phi,t) =  \phi_0(X) +  \delta \phi (X) Y_l^m (\theta,\phi) (t_0-t)^{-\sigma}, \nonumber 
\end{eqnarray}
where $V$ and $W$ are the angular velocities and $\delta V(X)$, $\delta W(X)$ are the corresponding perturbations in the self-similar frame and $Y_l^m(\theta,\phi)$ are the usual spherical harmonics. As in the study of HM99, the equations for $\delta V$ and $\delta W$ are identical so the perturbations are given the same amplitude, $\delta V(X)$. 
Replacing these expressions into Eqs.~(\ref{eq_hydro_ns}), we get 
%
\begin{eqnarray}
   (\sigma +2 + 3 \alpha_0) \delta R + (1 + \alpha_0) X \delta R ' + R_0  \delta U' + {2 R_0 \over X } \delta U    \nonumber \\
    -{l R_0 \over X} \delta V =  0,  \label{eq_pert_ns_1} \\
   (\sigma + \alpha_0) \delta U + (1 + \alpha_0) X  \delta U ' = - { \delta R' \over R_0} -  \delta \phi '  , \label{eq_pert_ns_2} \\
   (\sigma + \alpha_0) \delta V + (1 + \alpha_0) X  \delta V ' = - { (l+1)  \over X } \left( {\delta R \over R_0} +  \delta \phi \right)\label{eq_pert_ns_3} \\
\delta \phi '' + {2 \delta \phi' \over X } - {l(l+1) \over X^2} \delta \phi = \delta R. \label{eq_pert_ns_4} 
\end{eqnarray}
%
Thus, the system of Eqs.~(\ref{eq_pert_ns_1})-(\ref{eq_pert_ns_4}) consists of three first-order ordinary differential equations and one second-order 
ordinary differential equation.  Like in the spherical case, we solve it using a standard Runge-Kutta 
method.

\subsection{Inner boundary and shock condition in the non-axisymmetric case}

The boundary conditions of Eqs.~(\ref{eq_pert_ns_1})-(\ref{eq_pert_ns_4})
at $X \simeq 0$ are given by
\begin{eqnarray}
\nonumber
\delta R &\rightarrow&  R_1 X ^l, \\
\delta U &\rightarrow&  U_1 X ^{l-1}, \label{bound_cond_ns1}
 \\
\delta V &\rightarrow&  V_1 X ^{l-1}, \nonumber \\
\delta \psi &\rightarrow&  \psi_1 X ^l, \nonumber 
\end{eqnarray}
where
\begin{eqnarray}
 R_1 ={\rm cst} , \nonumber \\
(l+1) U_1 = l V_1, \label{bound_cond_ns2} \\
\phi _1 = - {R_1 \over R_0} - U_1 \left( {\sigma + \alpha_0 \over l} + (1+\alpha_0) {l-1 \over l} \right)
\nonumber
\end{eqnarray}
for the system to have a solution.
Eqs.~(\ref{eq_pert_ns_1})-(\ref{eq_pert_ns_4}) admit a family of solutions that 
resemble these boundary conditions and are given by Eqs.~(\ref{bound_cond_ns1}) and 
\begin{eqnarray}
R_1 = 0 , \nonumber \\
(l+1) U_1 = l V_1, \label{exact_sol_ns} \\
\phi_1 = {1 \over l} (\sigma + \alpha_0 + (1 + \alpha_0) (l-1) ) U_1. \nonumber
\end{eqnarray}
The situation here resembles what we found for the spherical problem: 
While these solutions are mathematically acceptable, they diverge 
at large $X$ so they do not represent physical solutions. 
Nonetheless, they will play an important role in finding a solution, as we will show later.

It is easy to see that this system also admits a critical point located at $X_{\rm crit} = 3$.
Again, the only solutions that satisfy the inner boundary conditions and cross the critical point 
are the trivial ones stated by Eqs.~(\ref{exact_sol_ns}), which for $l>2$ diverge at infinity.

Following the same line of thought, we look for solutions that present a shock at the vicinity 
of the critical point, $X_{\rm{shock}} = X_{\rm crit} + \delta X$ where $\delta X >0$.
The unperturbed solution given by 
$\delta R=0$, $\delta U=0$ is connected to a post-shock state given by the Rankine-Hugoniot conditions. 
The same applies to the other variables $\delta V$, $\delta \phi$ and $\delta \phi'$. They are 
all assumed to be zero in the inner part before the shock. These variables 
however are continuous so they are also equal to zero immediately after the shock.

The Rankine-Hugoniot conditions are identical to the spherical case.  The same 
calculations then lead to
\begin{eqnarray}
  \delta U Y^m _l (\theta,\phi) (t_0-t)^{-\sigma} = -  {2   \over 3} \delta X, \nonumber \\
  \delta R Y^m _l (\theta,\phi) (t_0-t)^{-\sigma}  =  {2 R_0 \delta X  \over 3} =  {4    \over 9} \delta X. 
\label{rankine_expan_ns}
\end{eqnarray} 
This implies that the surface of the shock itself is spherical only at the zeroth order. 
At the first order, the shock surface is described by a spherical harmonic, $Y ^m _l$.

\subsection{Limit at large radii for non-axisymmetric modes}
From Eqs.~(\ref{eq_pert_ns_1})-(\ref{eq_pert_ns_4}), it is easy to infer that the asymptotic 
behavior towards infinity is
\begin{eqnarray}
\delta R &\rightarrow& R_\infty X^{-v-1}, \nonumber \\
\delta U  &\rightarrow& U_\infty X^{-v}, \label{asym_ns}
 \\
\delta V  &\rightarrow& V_\infty X^{-v}, \nonumber \\
\delta \phi  &\rightarrow& \phi_\infty X^{-v+1}. \nonumber
\end{eqnarray}
By inserting these expressions into Eqs.~(\ref{eq_pert_ns_1})-(\ref{eq_pert_ns_4})
we obtain a fourth-order polynomial whose four roots are:
\begin{eqnarray}
   v &=& 3 \sigma - 4, \nonumber \\
   v &=& 3 \sigma + 1, \\
   v &=& - l +1, \nonumber \\
   v &=&  l + 2. \nonumber
\label{deter_ns}
\end{eqnarray}
The two first roots are the same ones we obtained before. 
The third branch, $v=-l+1$, leads to a strong divergence with $l$, with the exception of
the value $l=2$ for which it leads to the same asymptotic behavior as the solution
stated by Eqs.~(\ref{homol}).  
This asymptotic form is associated to the solution stated by Eqs.~(\ref{exact_sol_ns}) and it is, generally speaking, unphysical. 

However, if this branch is linearly combined with the solution ~(\ref{exact_sol_ns}) in such a way that their asymptotic behaviors compensate,
the new solution can be made to present a physically acceptable asymptotic behavior described 
by the root $v=3 \sigma -4$, like in the spherical case. 
It is again required that $v > -1$, so we must have
$\sigma > 1$.  Also, the condition that the velocity go to 0 at large $X$ implies again that $\sigma > 4/3$.

To summarize, in order to get a physically meaningful perturbations of the exact 
solution stated by Eqs.~(\ref{homol}), we need to combine 
the solution stated by Eqs.~(\ref{exact_sol_ns}) with the 
solution obtained by applying Rankine-Hugoniot conditions at the vicinity of the critical point
in such a way that the linear combination does not diverge too strongly.  

If $S_l$ is the desired solution and
$\phi_1$ is the value of the linear combination in Eqs.~(\ref{exact_sol_ns}), then
before the shock the solution is given by Eqs.~(\ref{bound_cond_ns1}) and Eqs.~(\ref{exact_sol_ns}). At the shock position 
the solution must satisfy the jump relations stated by Eqs.~(\ref{rankine_expan_ns}).
The difference with the spherical case is that here $u_{r,1}$ is 
not zero, but given by Eqs.~(\ref{exact_sol_ns}). 
This leads to
\small
\begin{eqnarray}
  \delta U  Y^m _l  (t_0-t)^{-\sigma} &=& -  {2   \over 3} \delta X - U_1 X_{shock} ^{l-1} Y^m _l  (t_0-t)^{-\sigma}, \nonumber \\
  \delta R Y^m _l  (t_0-t)^{-\sigma}  &=&  2 R_0 \left( {\delta X  \over 3} + U_1 X_{shock} ^{l-1} Y^m _l  (t_0-t)^{-\sigma} \right) . 
\label{rankine_expan_ns2}
\end{eqnarray} 
\normalsize
while 
\begin{eqnarray}
  \delta V   &=& V_1 X_{shock} ^{l-1}, \nonumber \\
  \delta \phi   &=&  \phi_1 X_{shock} ^{l+1} Y^m _l. 
\label{comp_expan_ns}
\end{eqnarray} 
Thus $S_l$ is a linear combination of the two above solutions. This is made clearer by writing
\begin{eqnarray}
  \delta U  Y^m _l  (t_0-t)^{-\sigma} &=& -  {2   \over 3} \delta X' +   U_1 X_{shock} ^{l-1} Y^m _l  (t_0-t)^{-\sigma}, \nonumber \\
  \delta R Y^m _l  (t_0-t)^{-\sigma}  &=&  {2 R_0 \over 3}  \delta X'.
\label{rankine_expan_ns2}
\end{eqnarray} 
where $\delta X' = \delta X + U_1 X_{shock} ^{l-1} Y^m _l  (t_0-t)^{\sigma}$. 
We see that the first terms of the right-hand side of Eqs.~(\ref{rankine_expan_ns2}) are identical to Eqs.~(\ref{rankine_expan_ns}),
while the second terms are compatible with Eqs.~(\ref{exact_sol_ns}). 

It is interesting that, unlike in the spherically symmetric problem, here the perturbations cannot vanish in the subsonic region. 
Since gravity is a non-local force, these perturbations are induced inside the subsonic region 
by the non-axisymmetric density distribution of the supersonic region.

Strictly speaking, since the spherical harmonics $Y^l_m$ also take 
negative values, there are locations where expressions (\ref{rankine_expan_ns}) and (\ref{rankine_expan_ns2}) place the shock in the subsonic regions of the core.  
In order for the shock to always be outside the critical point or, in other words, in order for $\delta X$ to always be positive,
these expressions should contain a spherically symmetric term.  This can be achieved by combining the non-axisymmetric perturbations with the spherically symmetric perturbation described by Eq.~(\ref{exact_sol}) for $\sigma=1$, so essentially adding an extra term, constant in ($\theta$,$\phi$), in Equations (\ref{rankine_expan_ns})-(\ref{rankine_expan_ns2}) that describe the Rankine-Hugoniot conditions.  We have omitted such a term for the sake of simplicity, but the subsequent analysis would remain identical.  Indeed, we have verified that the solutions depend only weakly on the inner boundary conditions.

\subsection{Results for the non-axisymmetric modes}

As in the spherical case, here as well we specify a position 
for the shock $\delta X$ small enough to ensure linearity. 
We use a step $d X=10^{-3}$ and integrate up to $X=10^4$.   
At the high $X$ limit we divide the potential by $X^{l+1}$, which yields the parameter $\phi_1$ (see Eq.~\ref{comp_expan_ns}). 
To obtain $S_l$ then we subtract the unphysical solution~(\ref{exact_sol_ns}) from the result of the numerical integration and 
recover the expected asymptotic behavior for values of $X$ smaller than $10^4$. 
An artifact of this subtraction is that, when $X \simeq 10^4$ the potential goes abruptly to zero.
The integration to larger values of $X$ compared to the spherical problem
enables us to discard the diverging term while maintaining a physical behavior for a large enough range of $X$ values. 

\setlength{\unitlength}{1cm}
   \begin{figure} 
   \begin{picture} (0,18)
      \put(0,13.5){\includegraphics[width=9cm]{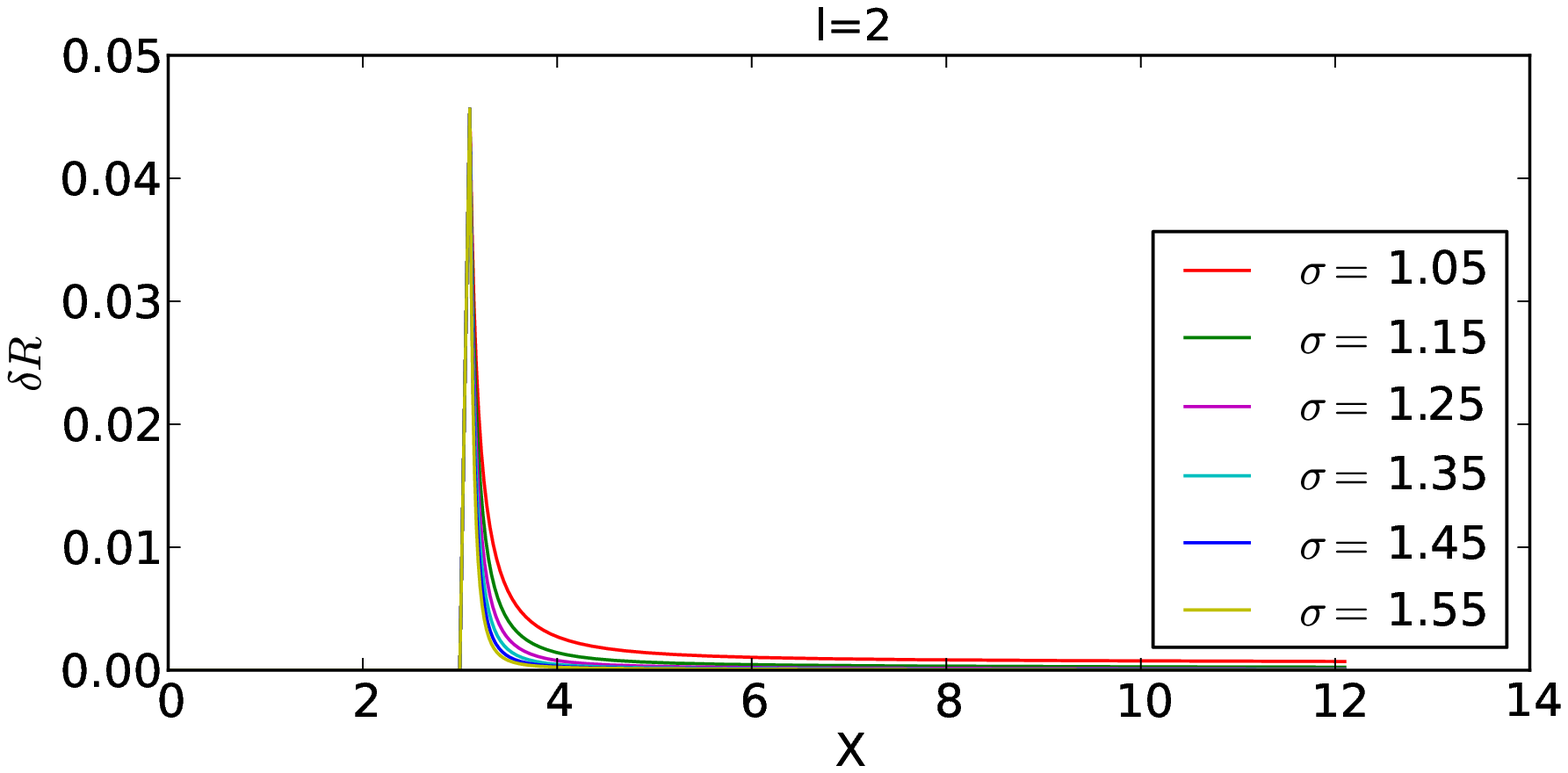}}
      \put(0,9){\includegraphics[width=9cm]{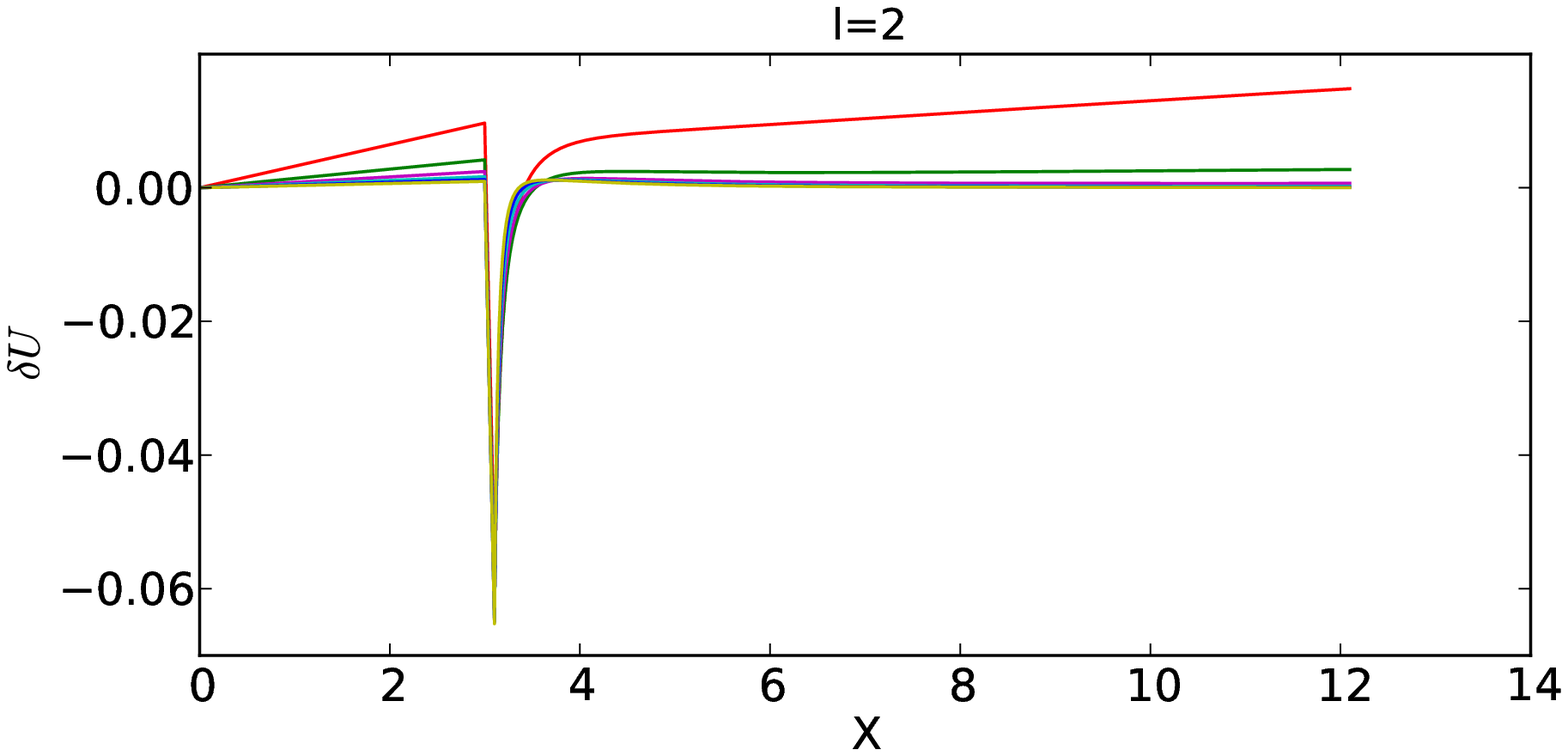}}
      \put(0,4.5){\includegraphics[width=9cm]{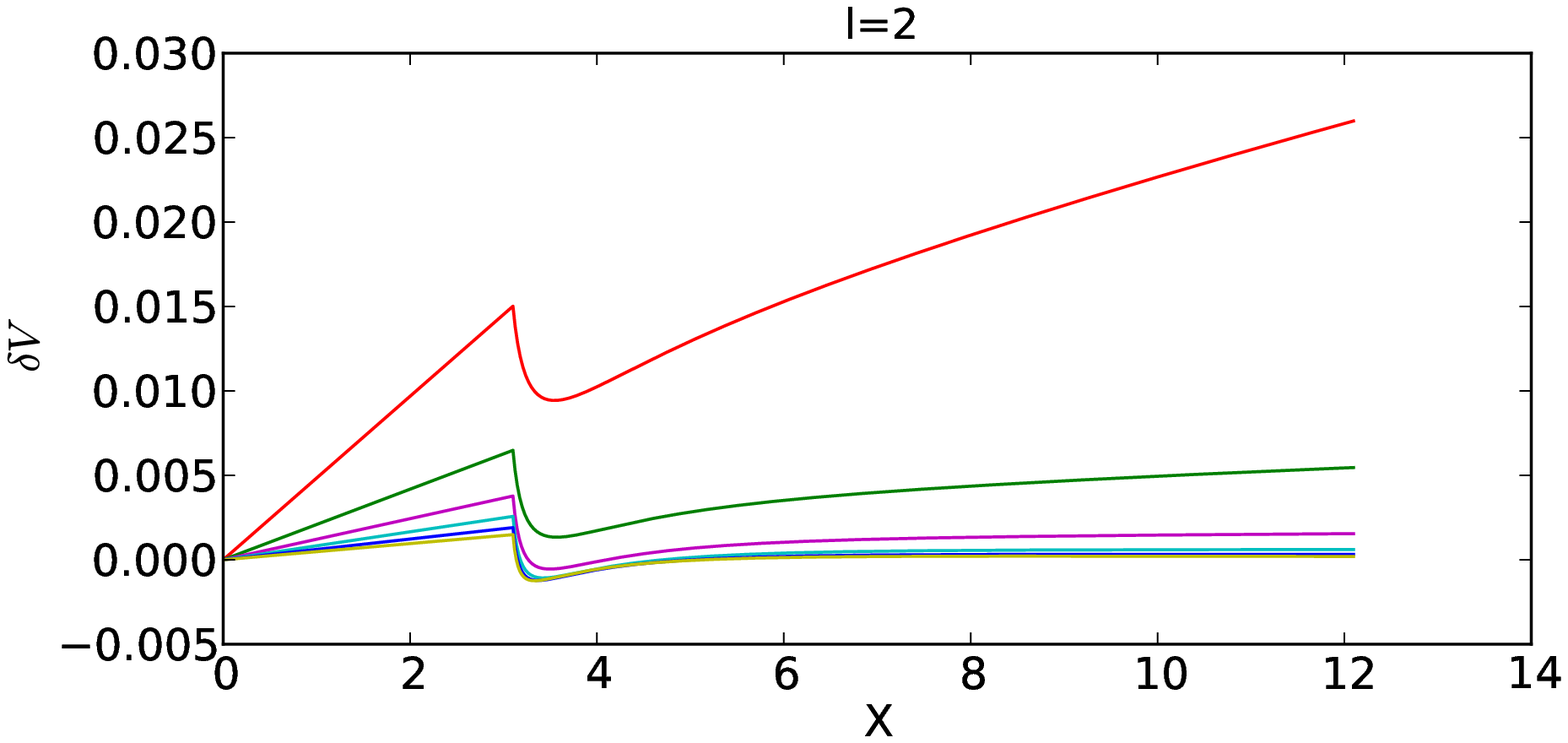}}
      \put(0,0){\includegraphics[width=9cm]{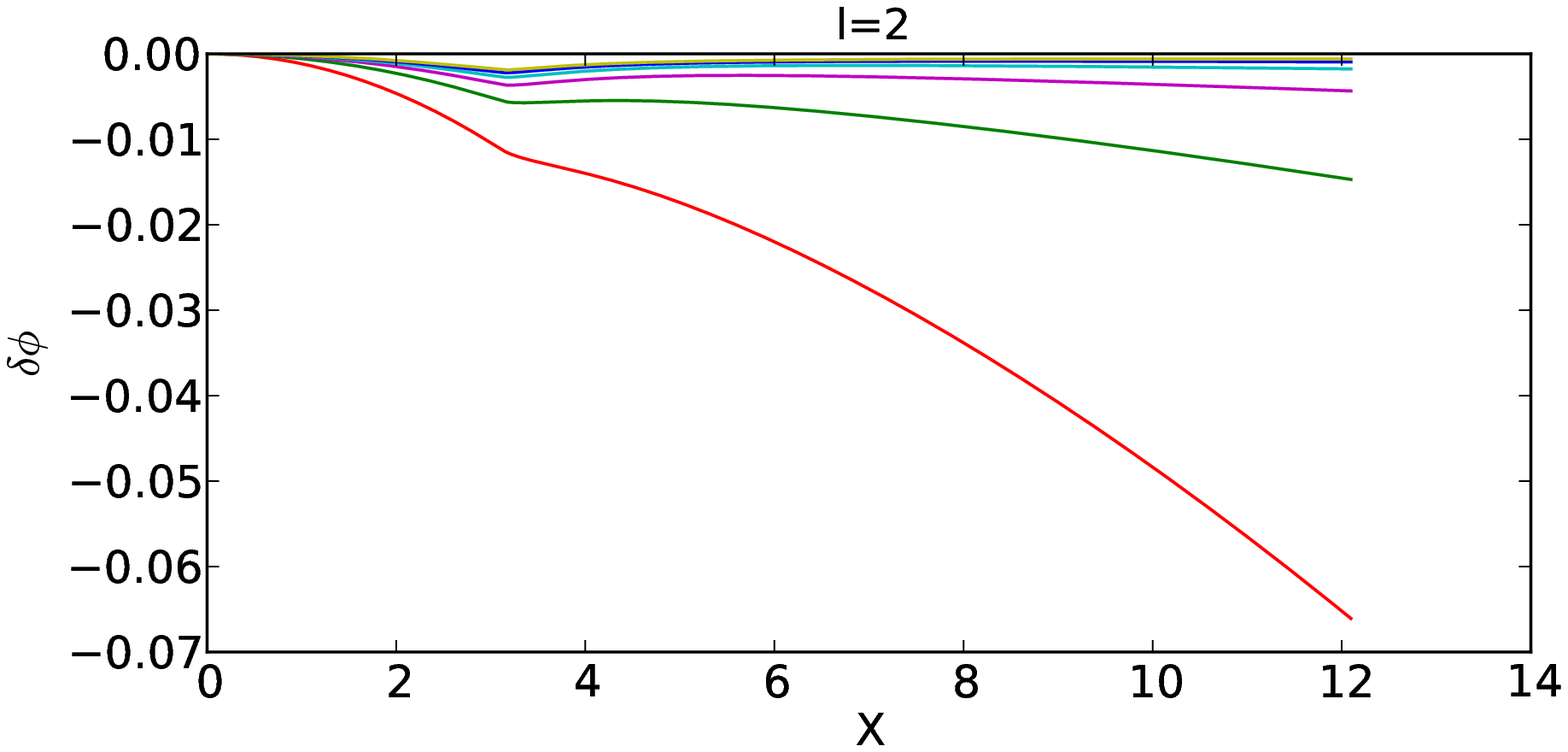}}
   \end{picture}
   \caption{From top to bottom, density $R$, radial velocity $U$, azimuthal 
velocity $V$ and gravitational potential $\phi$ for the 
 $l=2$ mode and for a series of growth rates.}
   \label{l2}
\end{figure}

\setlength{\unitlength}{1cm}
\begin{figure} 
   \begin{picture} (0,18)
      \put(0,13.5){\includegraphics[width=9cm]{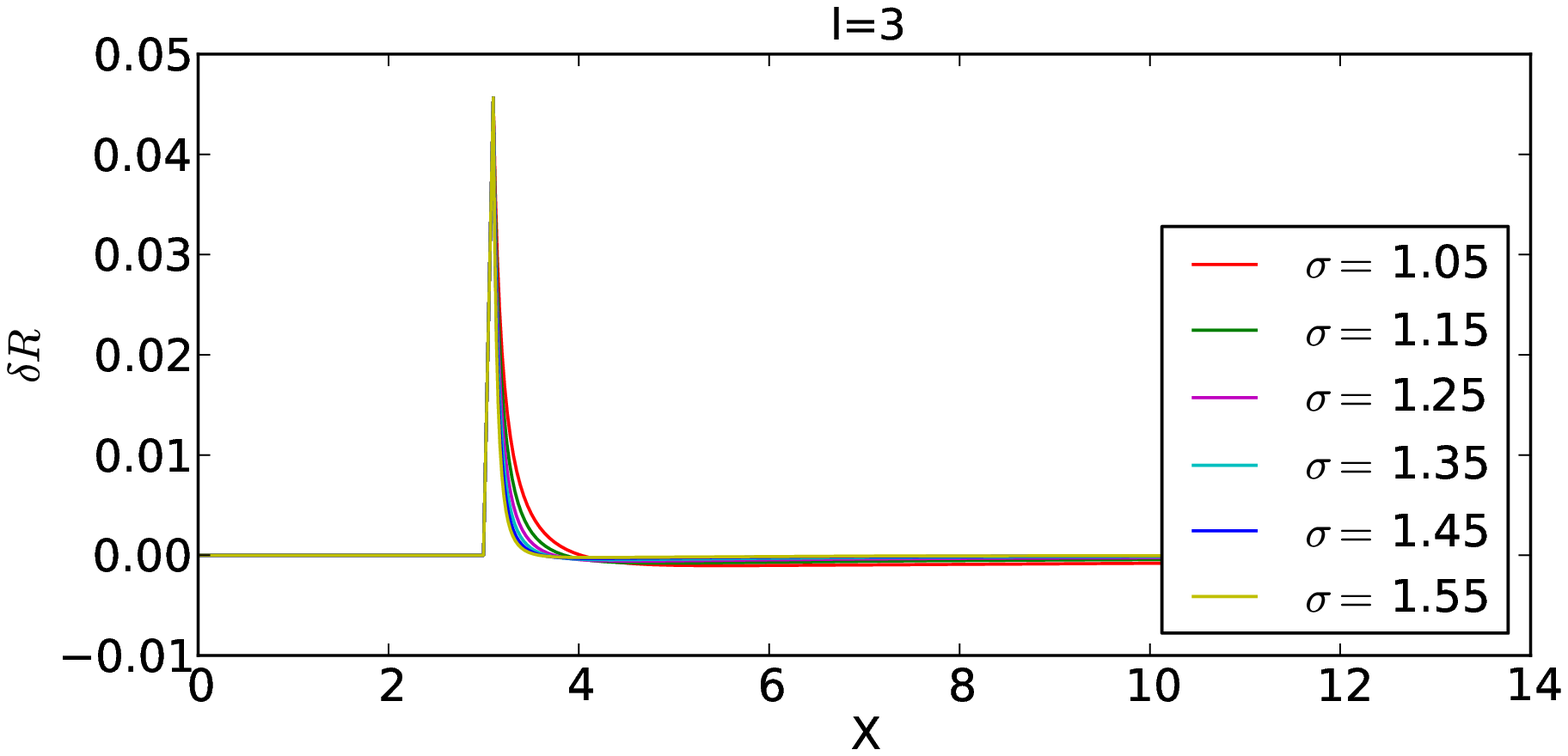}}
      \put(0,9){\includegraphics[width=9cm]{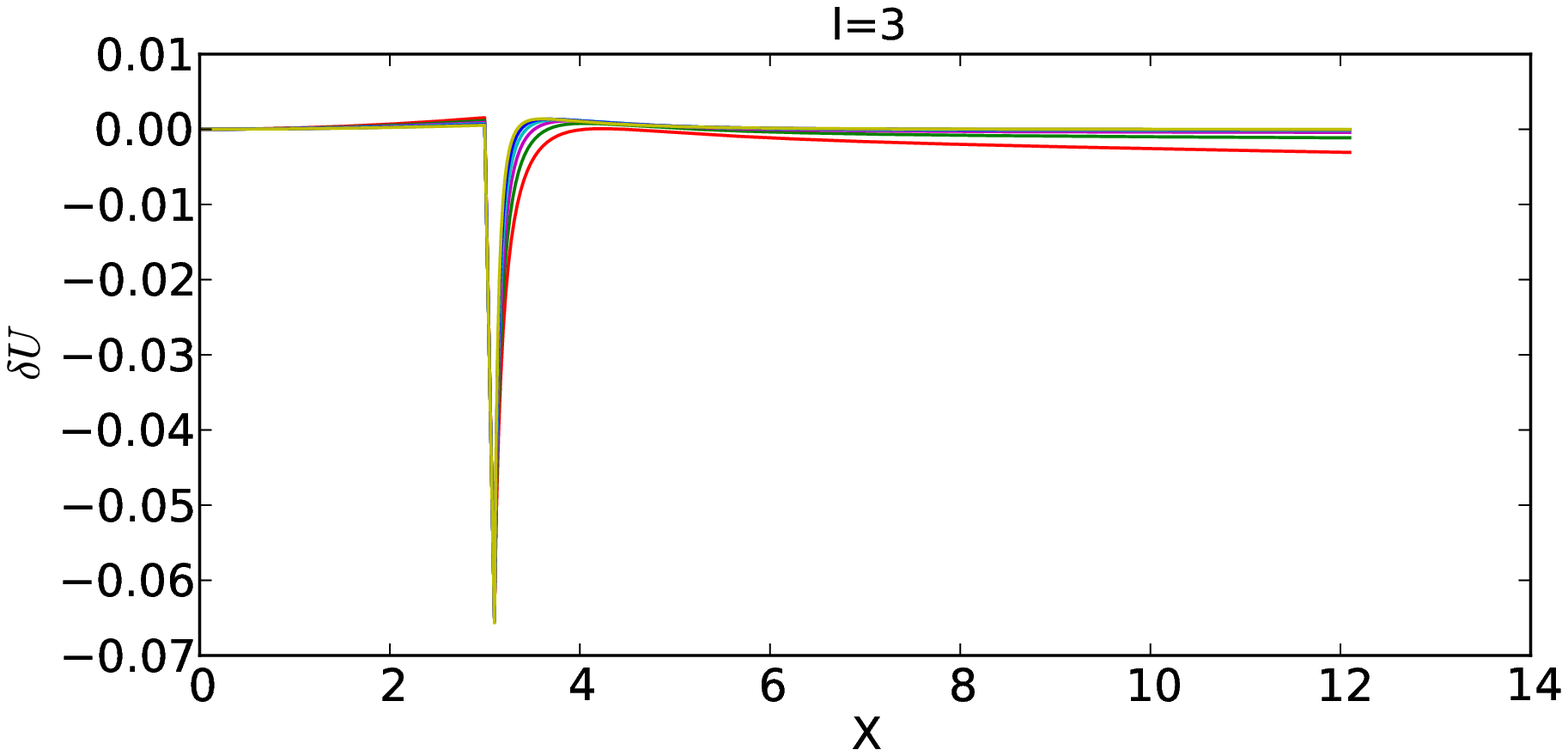}}
      \put(0,4.5){\includegraphics[width=9cm]{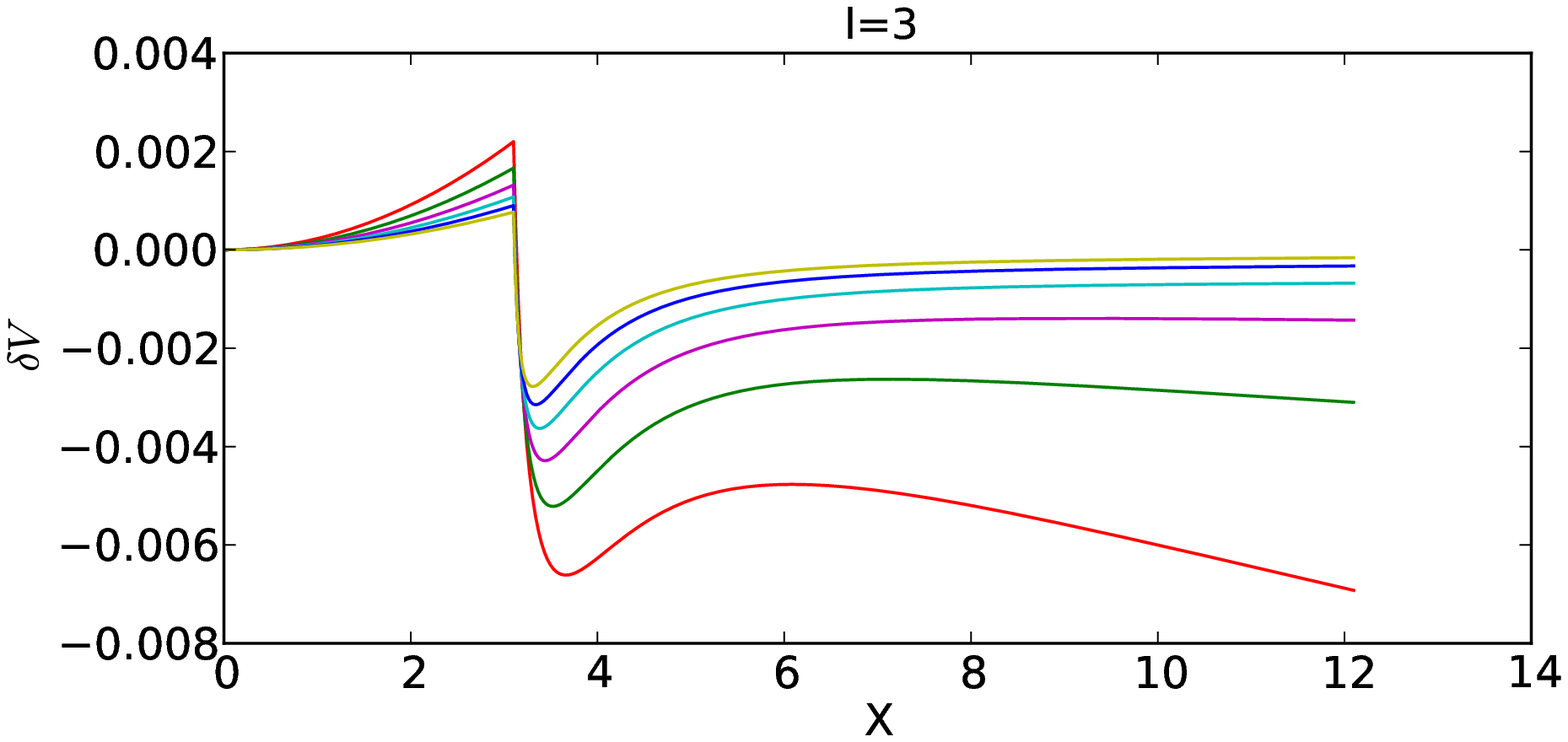}} 
      \put(0,0){\includegraphics[width=9cm]{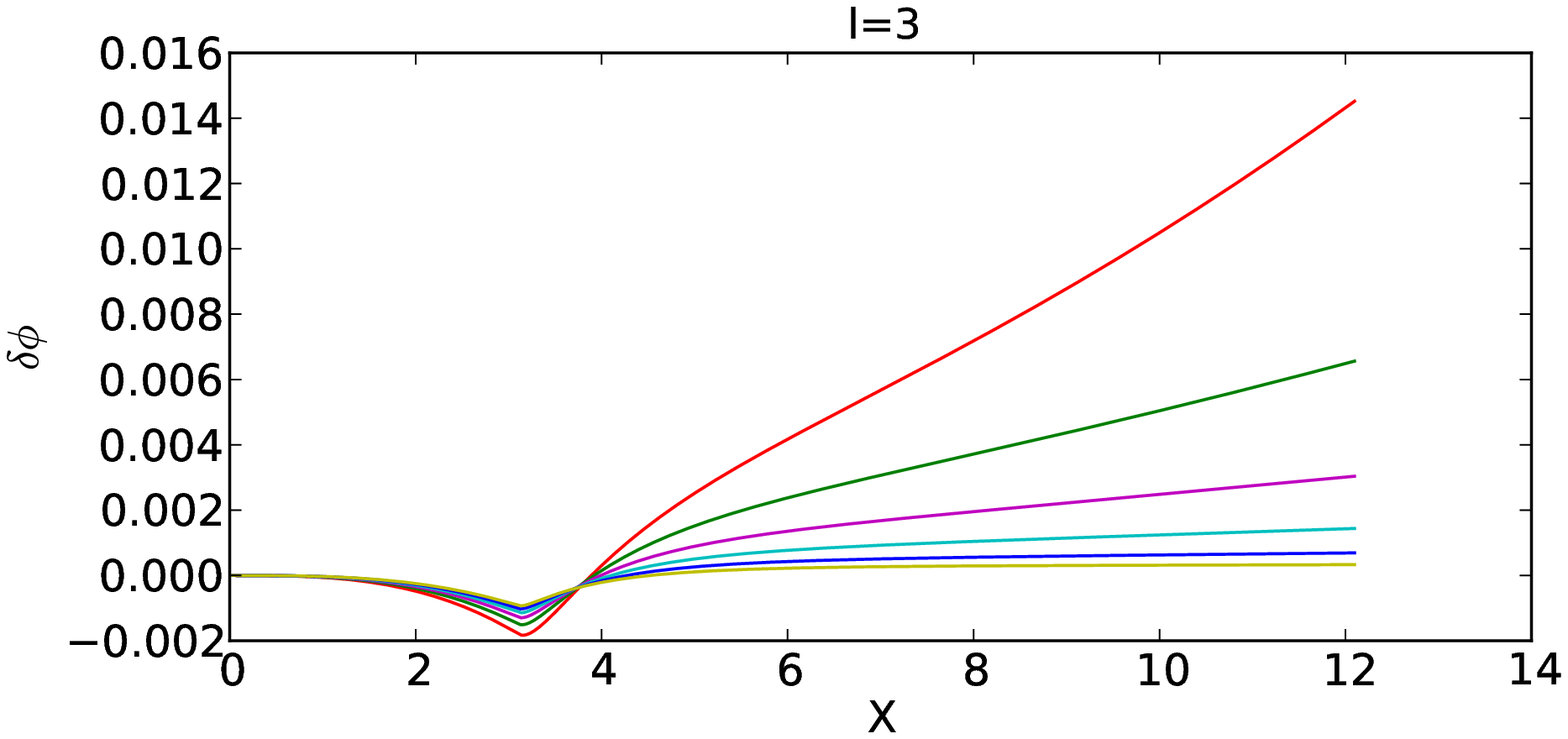}} 
    \end{picture}
   \caption{Same as Fig.~\ref{l2} for the  $l=3$ mode.}
\label{l3}
\end{figure}

\setlength{\unitlength}{1cm}
\begin{figure} 
   \begin{picture} (0,18)
      \put(0,13.5){\includegraphics[width=9cm]{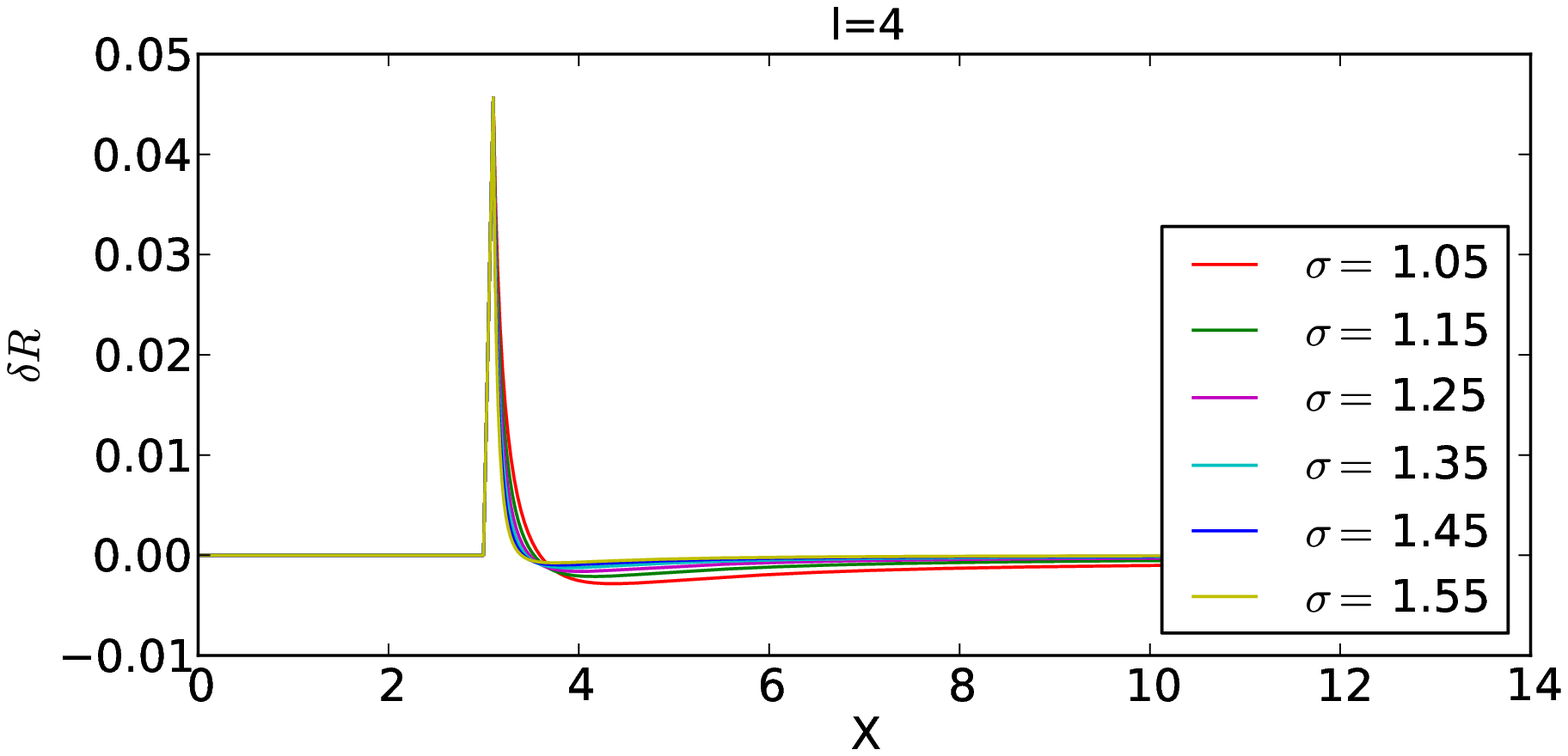}}
     \put(0,9){\includegraphics[width=9cm]{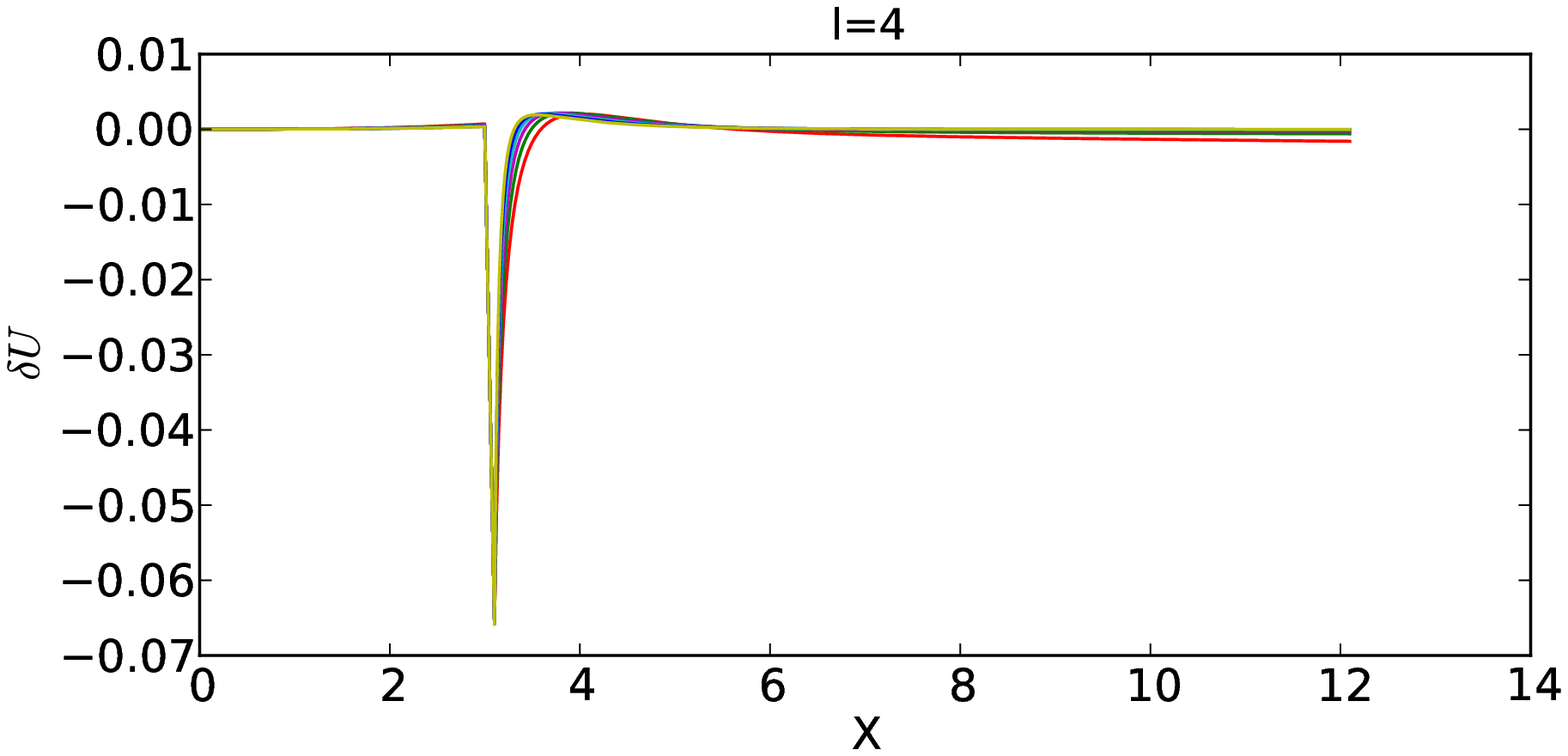}}
     \put(0,4.5){\includegraphics[width=9cm]{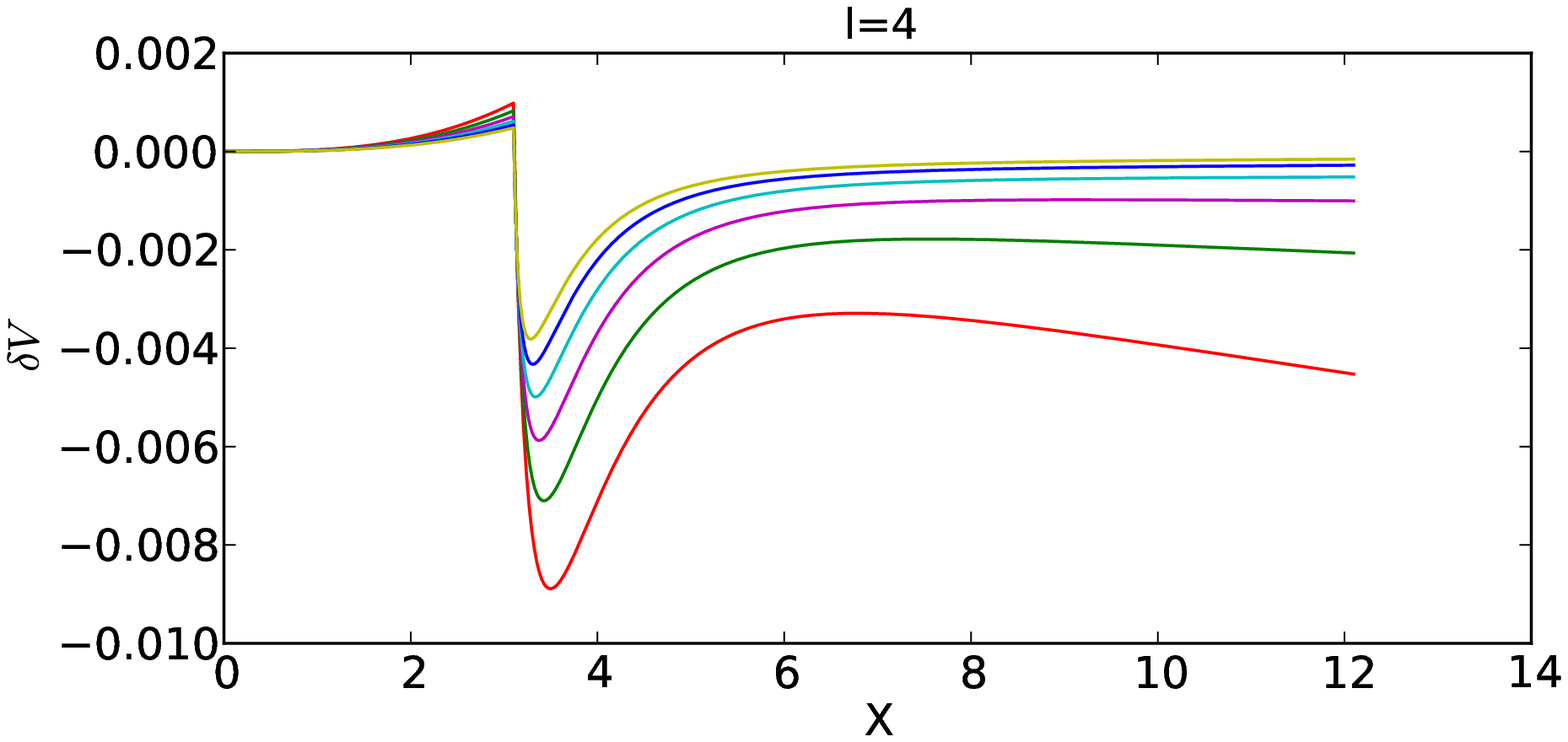}}
    \put(0,0){\includegraphics[width=9cm]{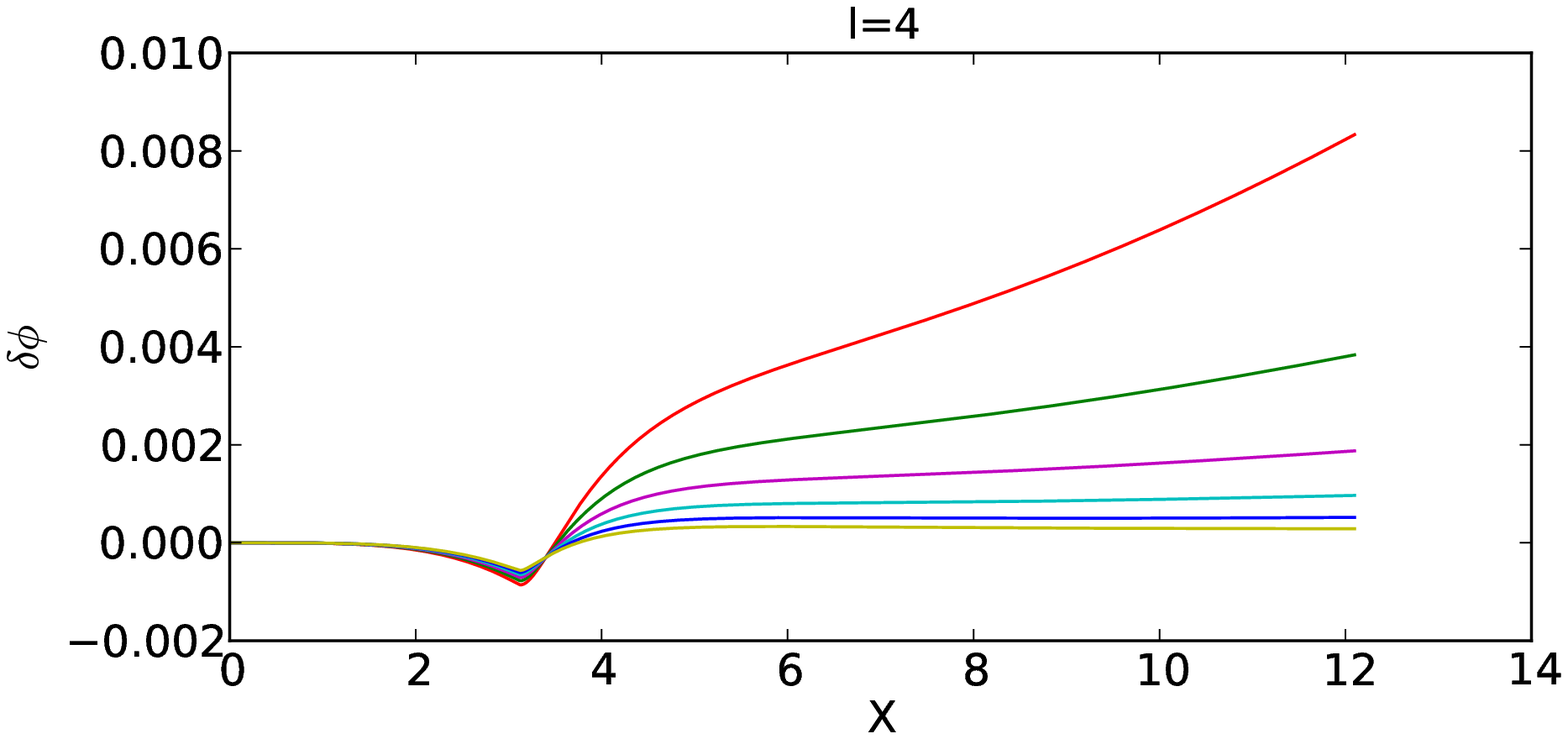}}
   \end{picture}
   \caption{Same as Fig.~\ref{l2} for the  $l=4$ mode.}
   \label{l4}
\end{figure}

Figures~\ref{l2},~\ref{l3} and~\ref{l4} show the density $R$, radial and azimuthal velocities $U$ and $V$ and the gravitational potential $\phi$ for the azimuthal wave numbers $l=2, \, 3$ and 4 respectively and for a series of growth rates $\sigma$.
These profiles show the solutions described by  Eqs.~(\ref{exact_sol_ns}) in the inner subsonic regions, then 
a shock in the neighborhood of the critical point $X_c \simeq 3$ 
and finally a behavior at large $X$ consistent with the asymptotic analysis above and similar to the spherically symmetric case. 

Although not identical, the three modes behave similarly to the spherically symmetric example.  
The values of the various fields tend to be smaller for larger $l$ because the gravitational potential has stiffer spatial variations in the azimuthal directions, which lowers the resulting force inside the subsonic regions.   
Like in the simple one-dimensional case, here as well higher values of $\sigma$ lead to sharply peaked profiles around the shock position.  
This appears to suggest that the physically relevant values of $\sigma$ are between 1 and $\simeq 1.5$, but at the same time, there does not seem to be an obvious way of deciding whether a particular growth rate is expected or preferred.  
For this reason and also for the sake of verifying the reliability of the present approach, we 
perform numerical simulations, which are presented in the following section. 

\section{Stability of collapsing cores in numerical simulations}

The above analysis is complemented with a series of numerical experiments,
consisting of collapsing cores with different initial ratios of gravitational to thermal energy (virial parameter $\alpha$), and different initial perturbations.
These simulations are a very efficient way to witness the behavior of a linearly perturbed core in the full complexity that three-dimensionality entails, meaning the
inclusion of the azimuthal and vertical velocity components.  It also allows us to probe the limit for fragmentation as the thermal support of the core increases.

\subsection{Code and Initial conditions}

The simulations are performed with the publicly available  Adaptive Mesh Refinement (AMR) code RAMSES \citep{Teyssier_02},
which solves the equations of hydrodynamics on a Cartesian grid with a second-order Godunov scheme and comes with wide range of options for the simulated physical processes.  In this case we take advantage of its AMR capabilities by requiring that the Jeans length be always resolved with at least 20 grid cells. An isothermal equation of state is used throughout.

The initial condition of each simulation is a core of uniform density $\rho_0$, equal to $4\cdot10^{-18}$ gr cm$^{-3},$
and uniform temperature T, equal to 10 K.  To this profile we add a linear perturbation, given by

\begin{equation}
   \delta\rho = \left[\epsilon_1 \cos{(k_r\cdot 2 \pi r/L_c)}\cdot(1+\epsilon_2 Y_l^m(\theta,\phi)\right]
\end{equation}
where $k_r=0.25$ is the radial wavenumber of the perturbation, $r$ is the distance from the core
center, $L_c$ is the core outer radius and $\epsilon_1$, $\epsilon_2$ the amplitudes of each component of the perturbation, which, according to the case, 
have values from $0.05$ to $0.1$.
The core resides in a uniform medium of pressure and density 100 times smaller than the core's interior.

For all these simulations we employ periodic boundary conditions; when self-gravity is involved, such boundaries can potentially affect the evolution of the system, so the core is placed at the center of the computational volume, at a distance of two core radii from each box side.  This was shown to be enough to avoid boundary effects by convergence tests we performed for the largest core.

\subsection{Stability as a function of cloud parameters}

In this study of the core stability, we vary only two parameters: the azimuthal wave number $l$ of the spherical harmonics and
the virial parameter of the core, which gives the thermal over gravitational energy ratio in the core and is defined as
\begin{equation}
   \alpha = \frac{15 c_s^2}{8\pi G\rho_0 L_c^2},
\end{equation}
where $c_s$ is the sound speed in the core interior.
Here the temperature of the core is held constant and $\alpha$ is varied by changing the radius $R$ of the core.
We have not varied the radial wavenumber of the perturbation, which means the $l=0$ case is a single shell around the center of the core.

\begin{figure}

    \includegraphics[width=0.496\linewidth]{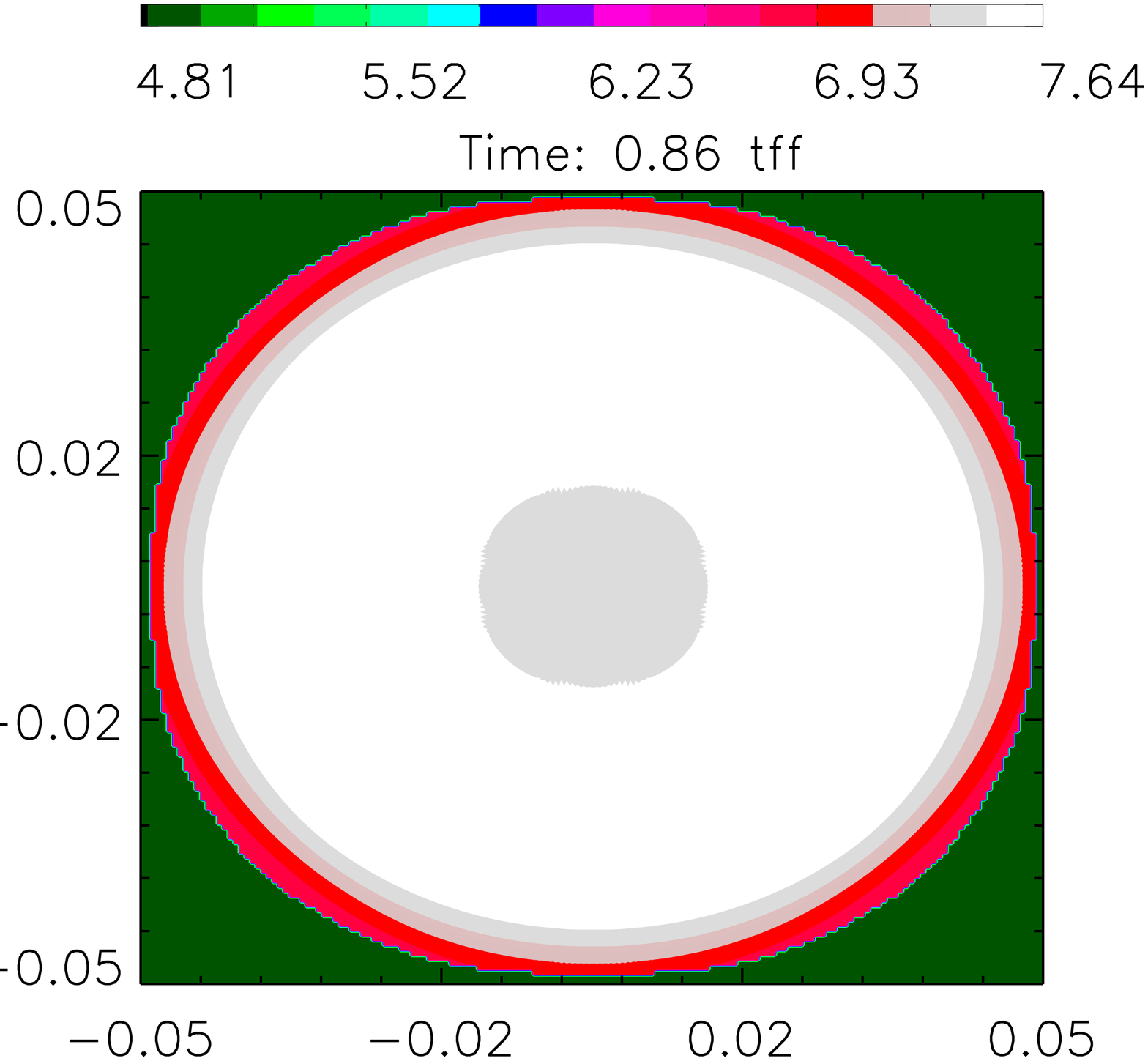}
    \includegraphics[width=0.496\linewidth]{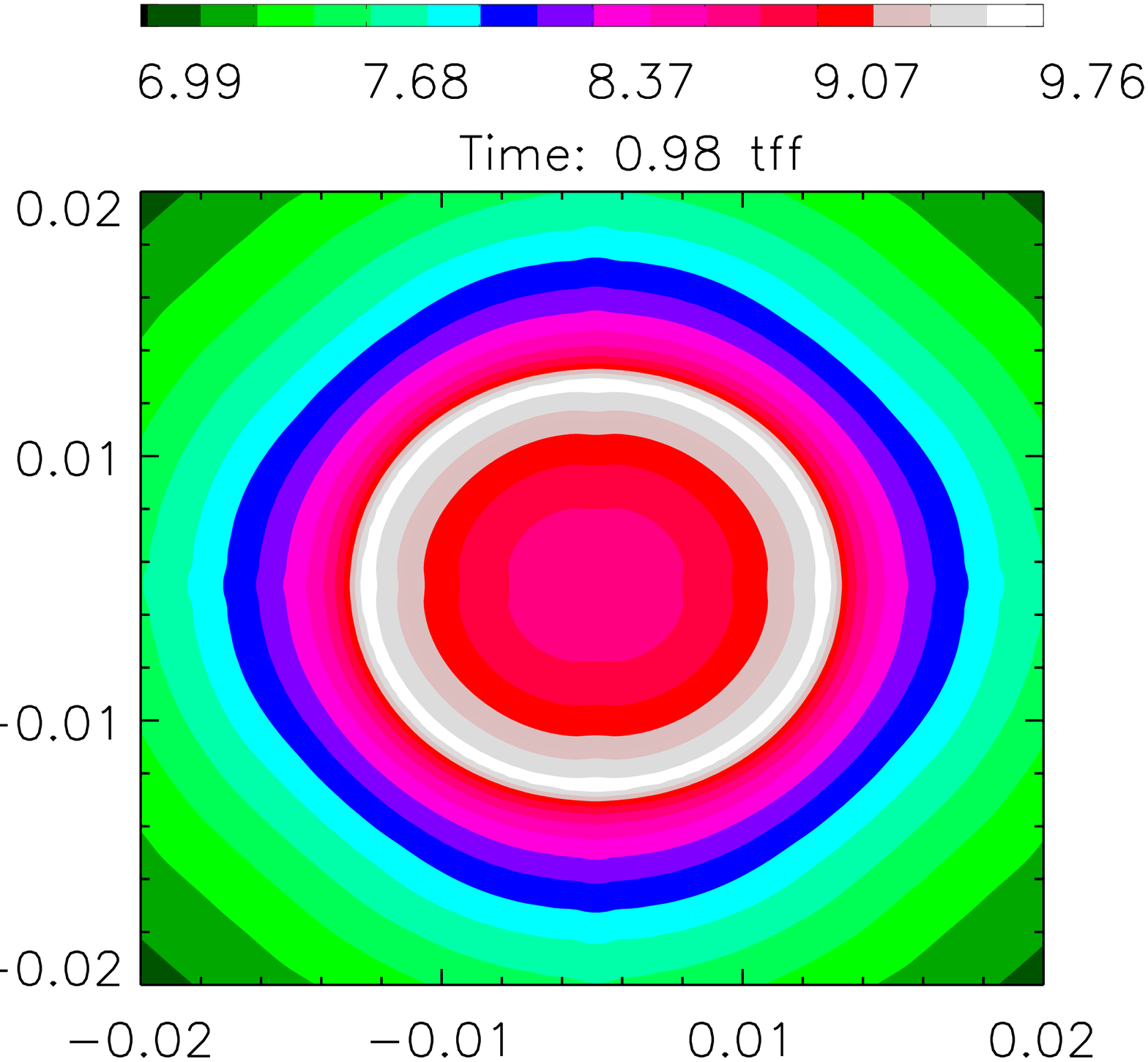}
   \includegraphics[width=0.496\linewidth]{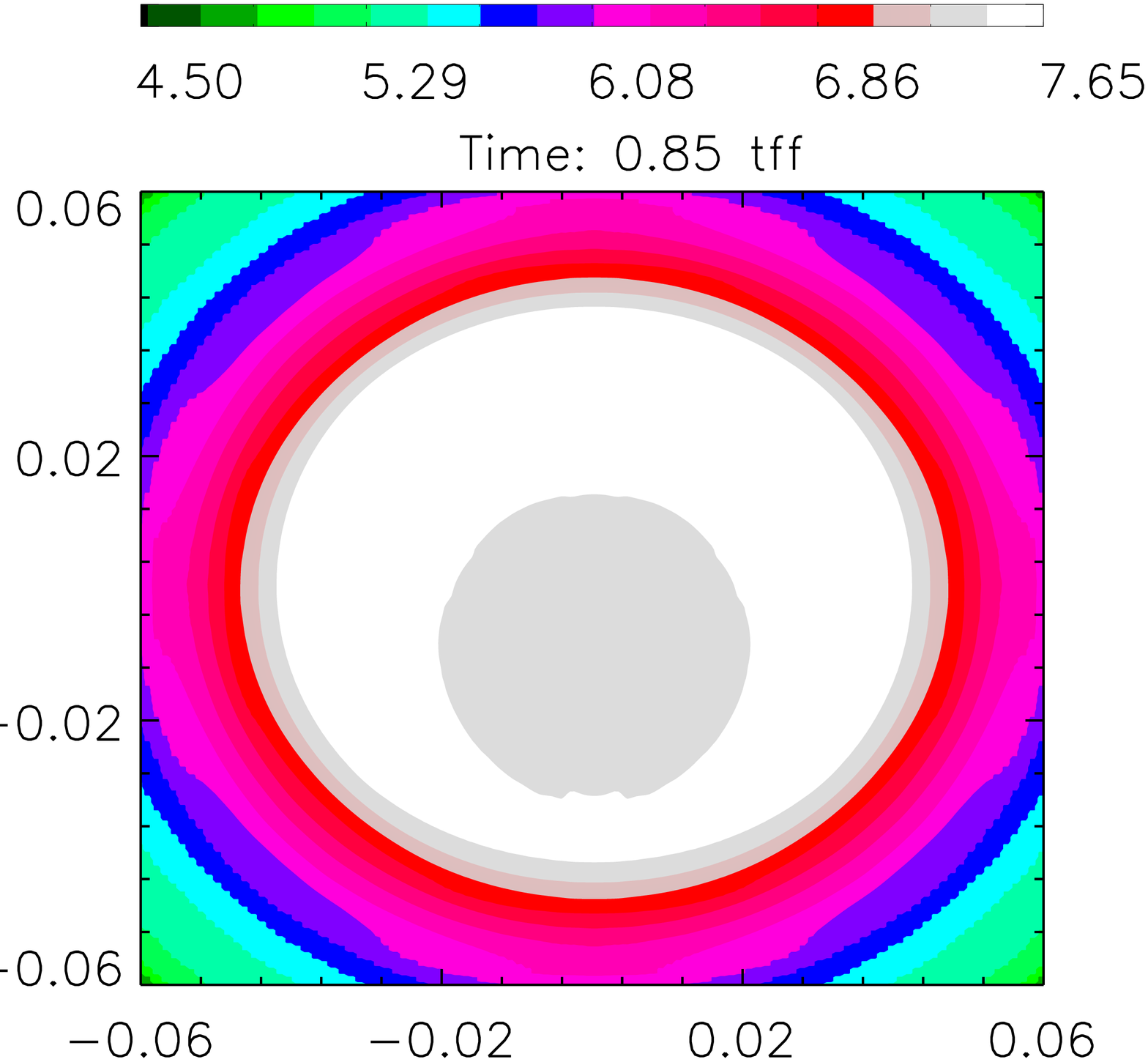}
  \includegraphics[width=0.496\linewidth]{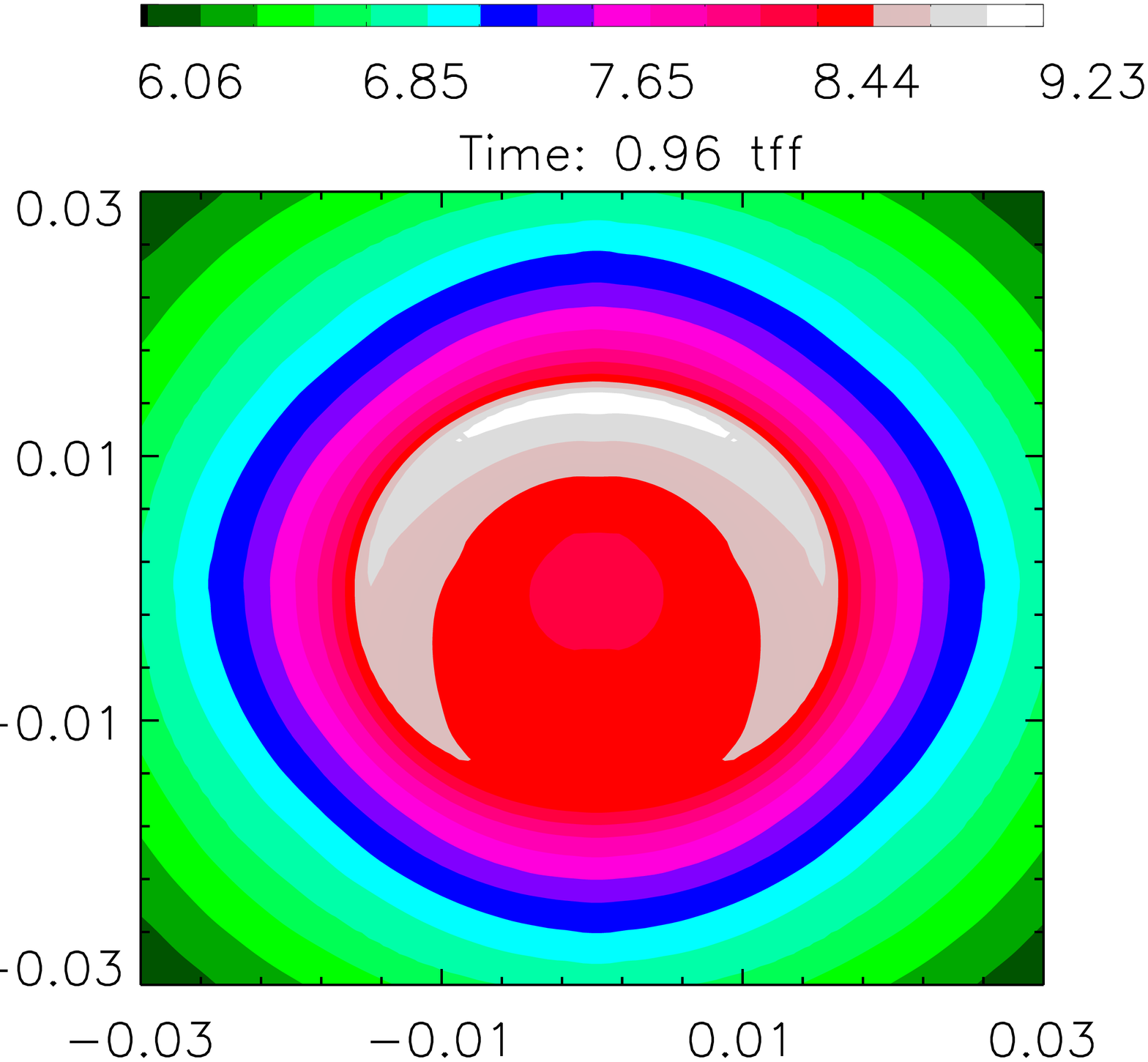}
    \includegraphics[width=0.496\linewidth]{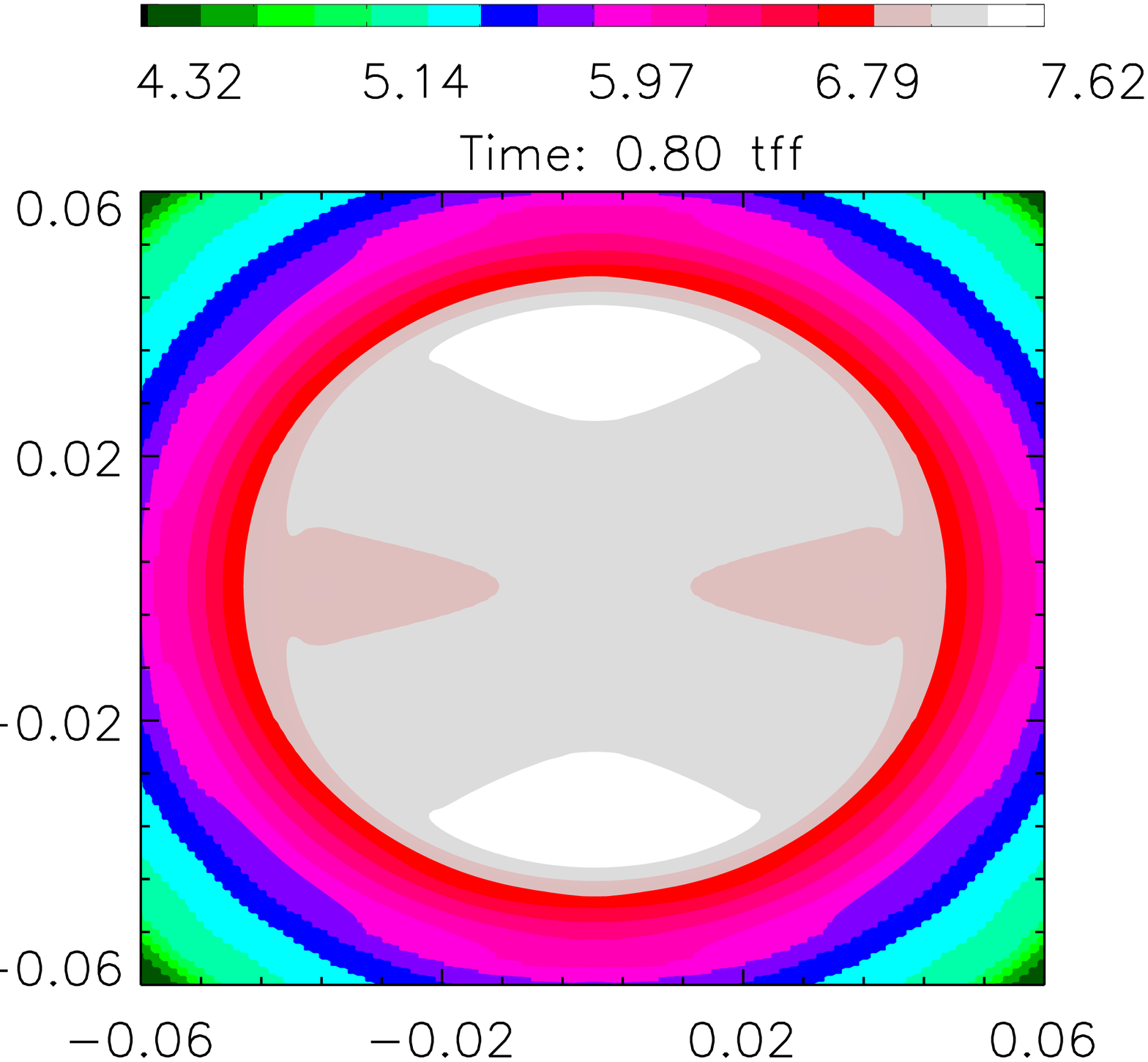}
    \includegraphics[width=0.496\linewidth]{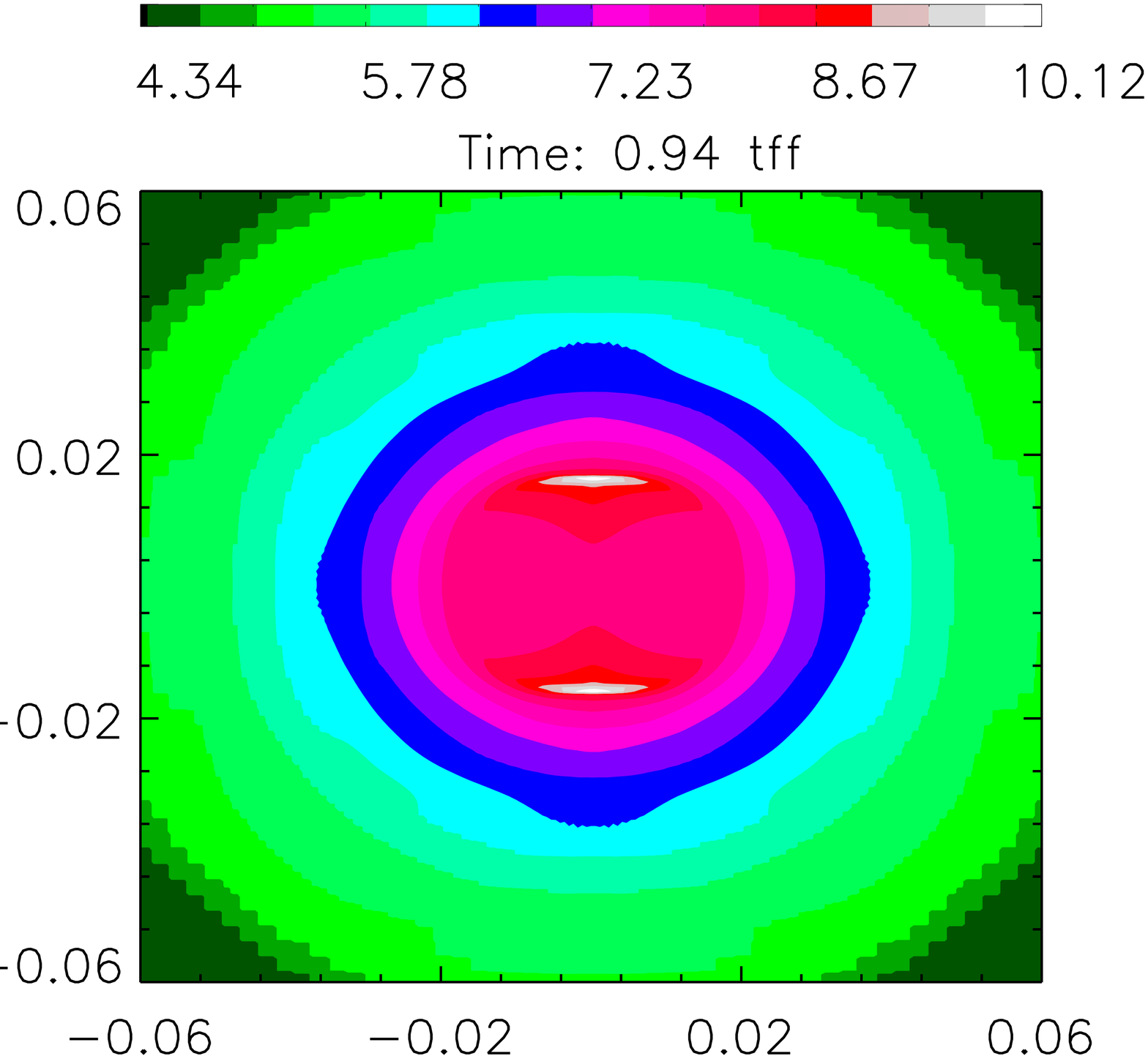}
    \includegraphics[width=0.496\linewidth]{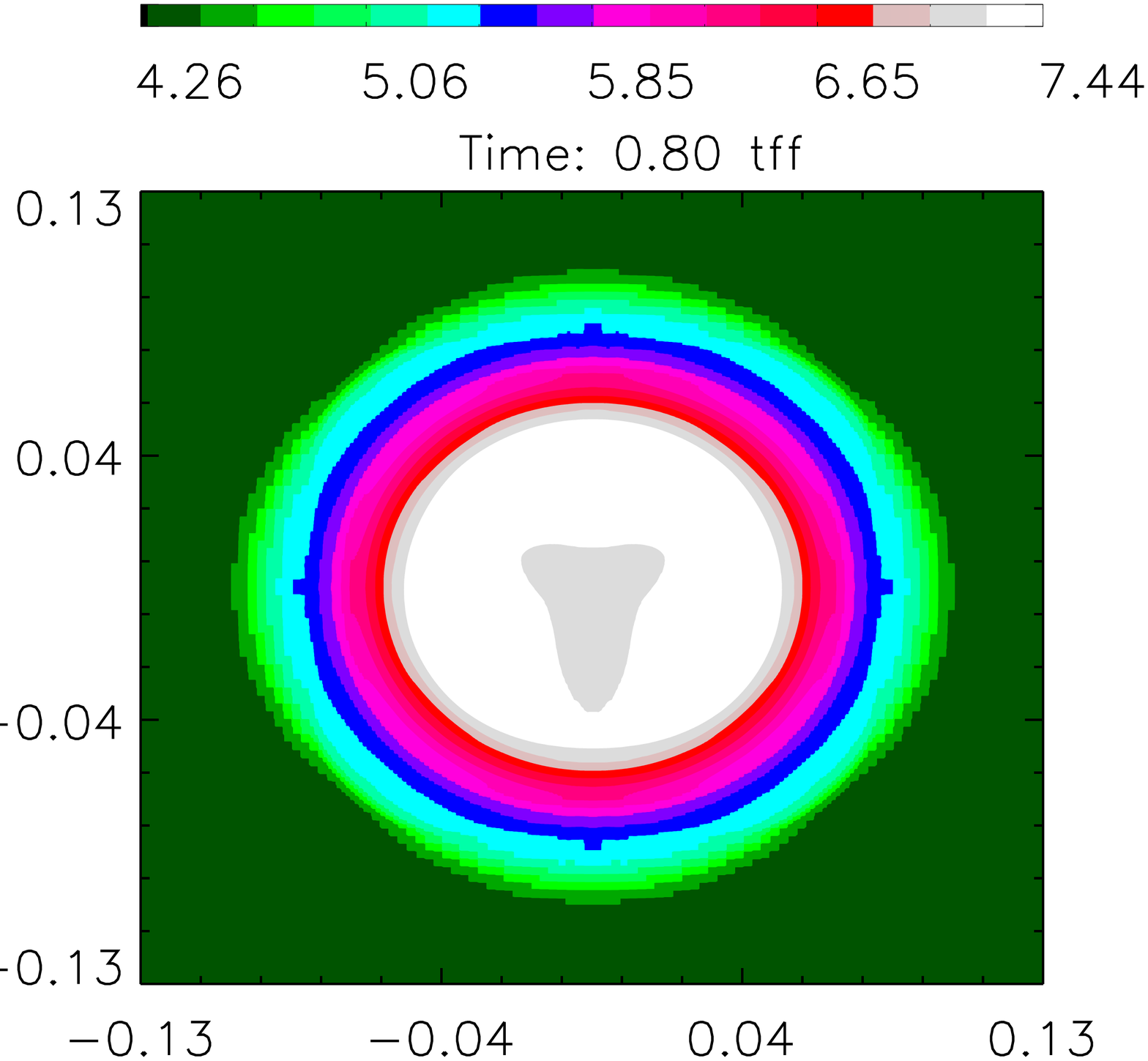}
    \includegraphics[width=0.496\linewidth]{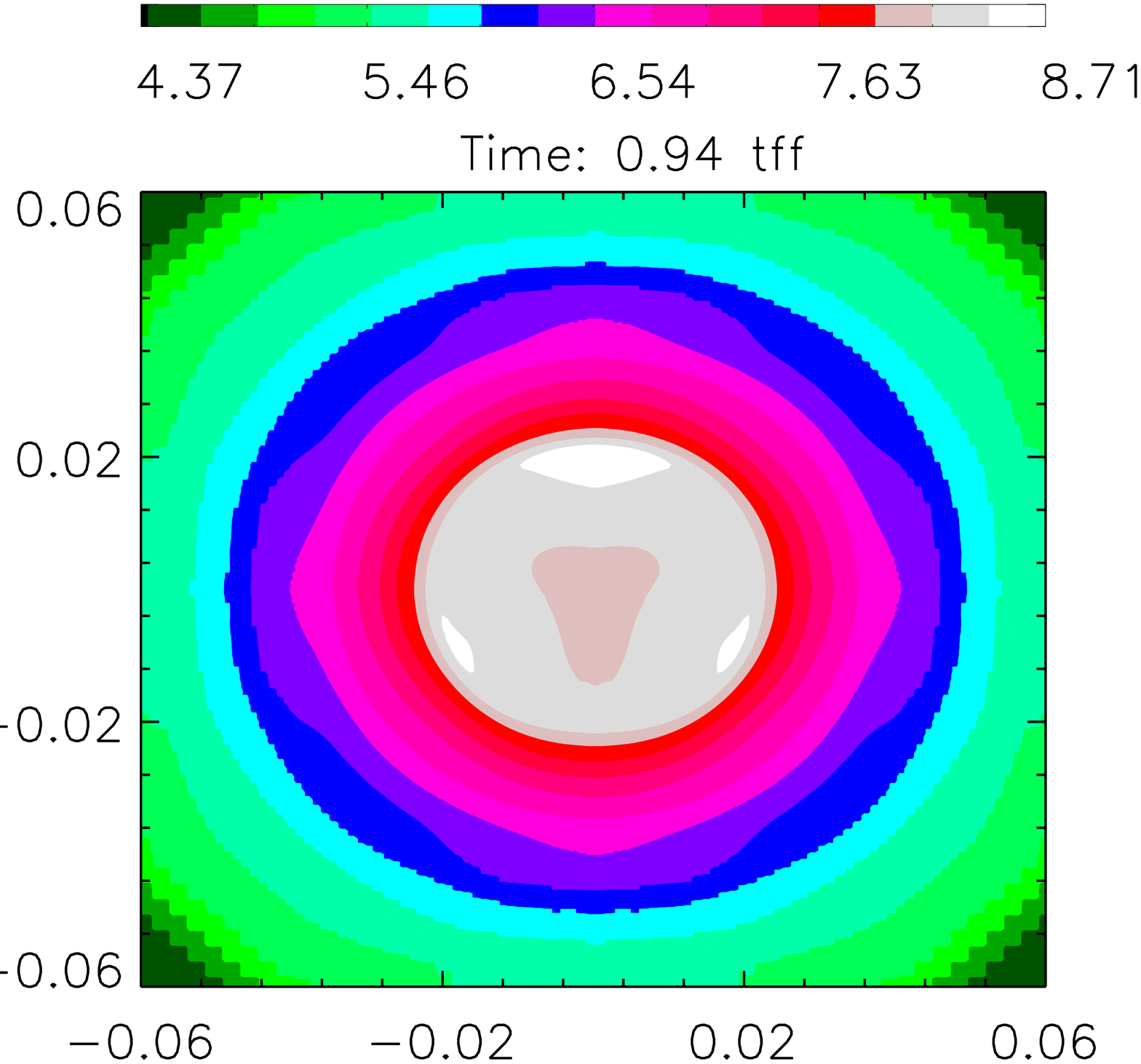}

    \caption{Contours of the log $n_H$ in units of $m_H$ cm$^{-3}$ along the yz plane.  From top to bottom: $l=0$, $l=1$, $l=2$ and $l=3$ modes for virial parameter $\alpha=0.006$.  The units along each axis are parsecs and $t_{ff}$ stands for free-fall time.}
     \label{a0.006}
\end{figure}

\begin{figure}
    \includegraphics[width=0.496\linewidth]{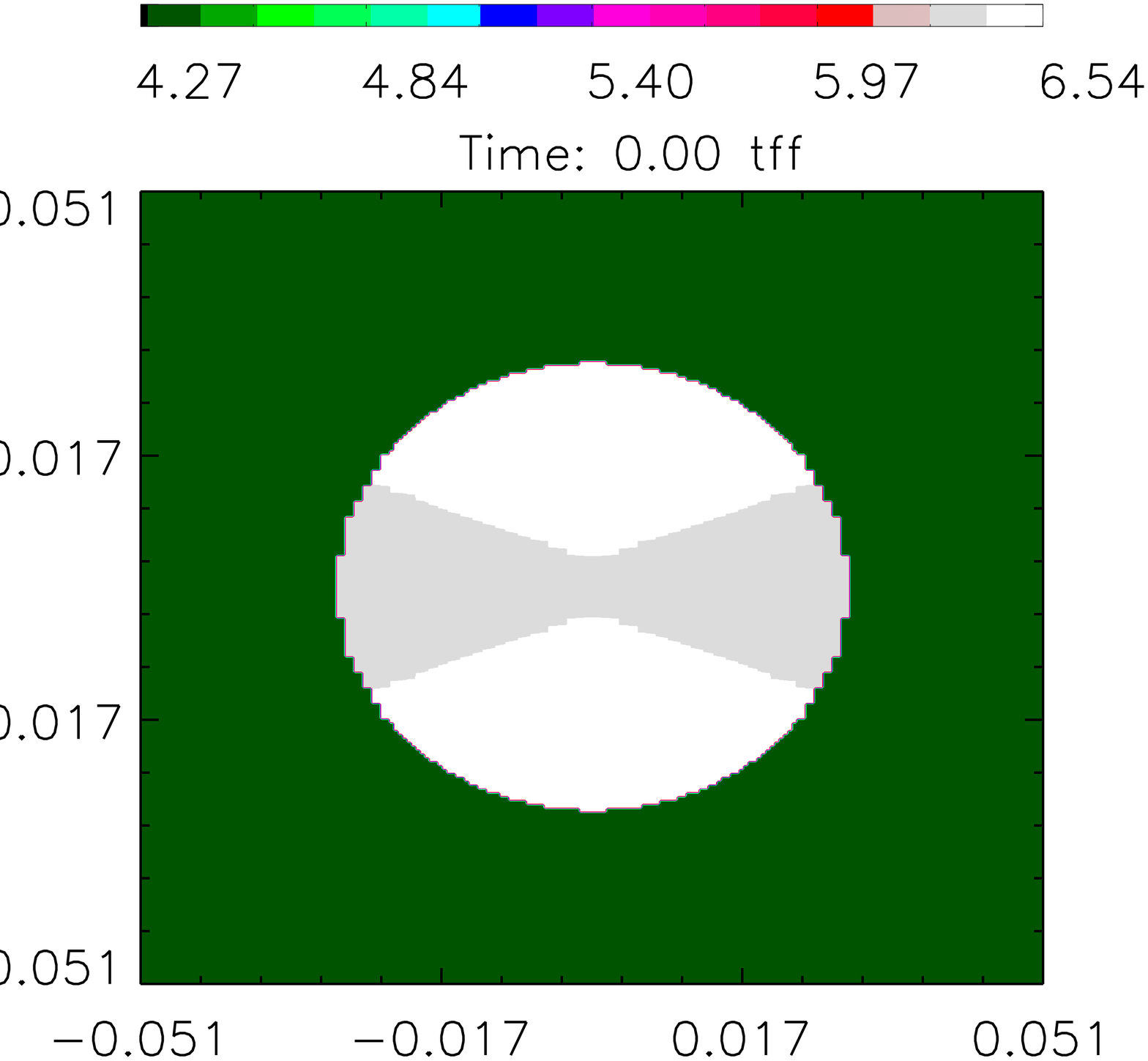}
    \includegraphics[width=0.496\linewidth]{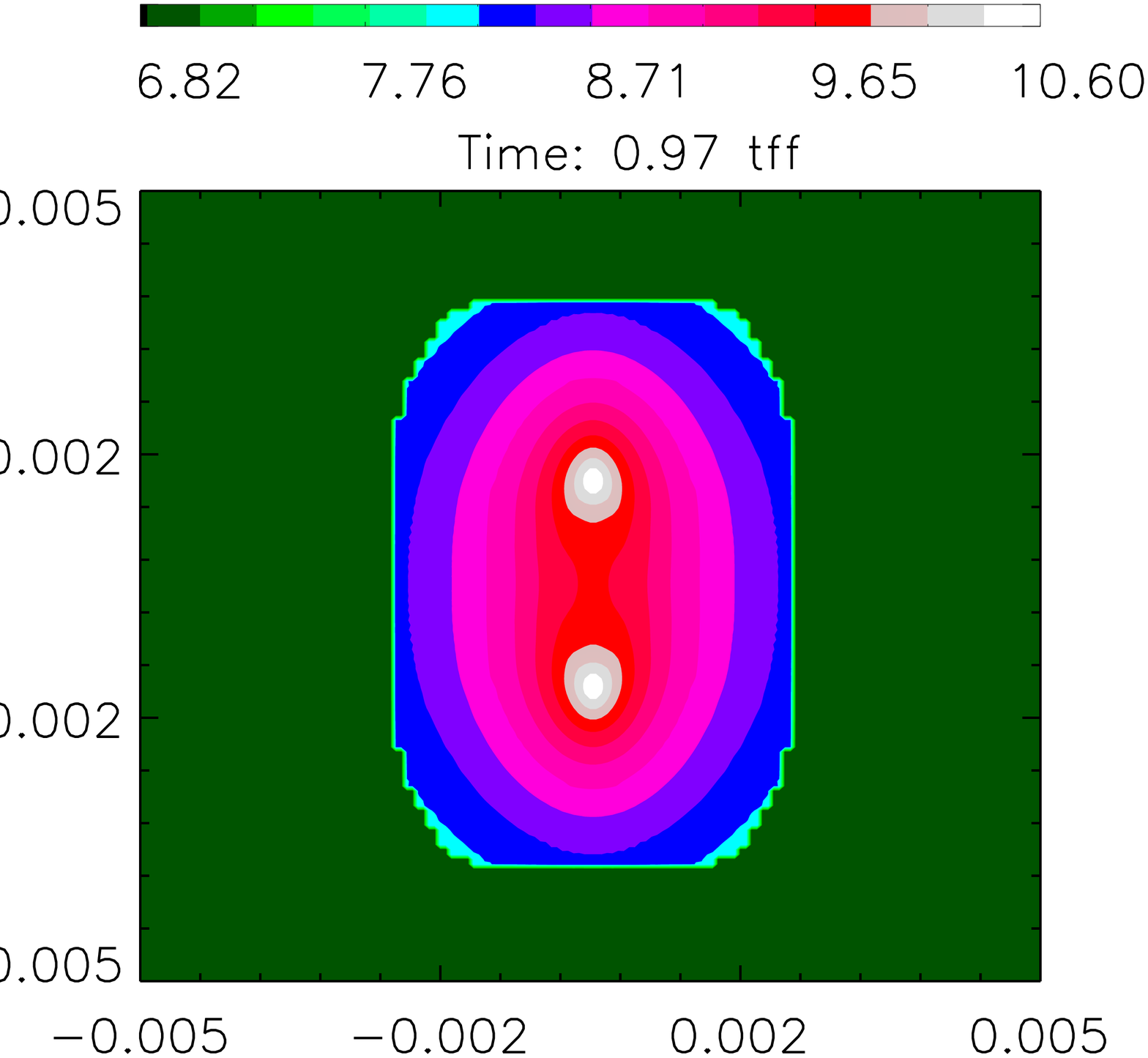}
    \includegraphics[width=0.496\linewidth]{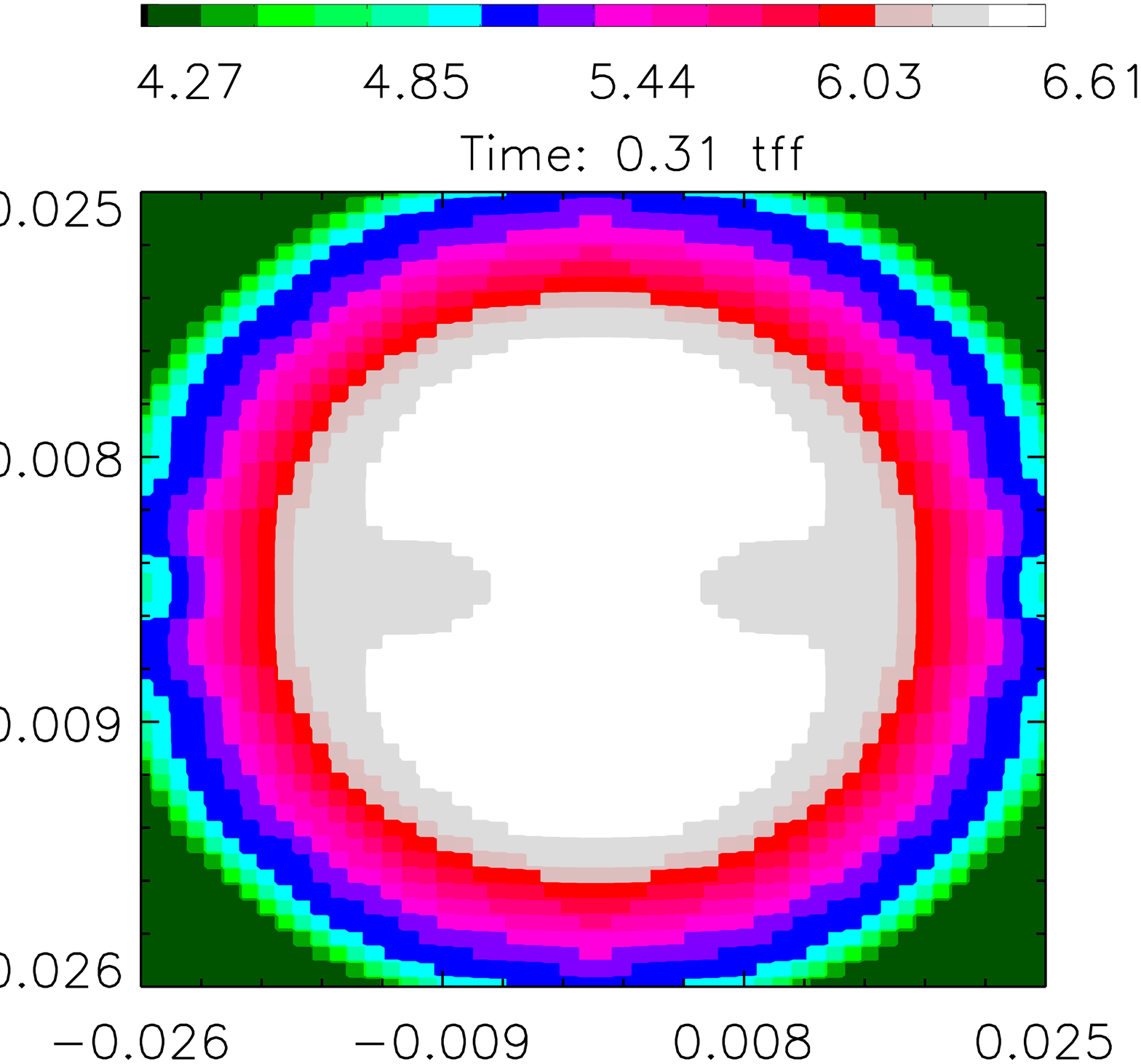}
    \includegraphics[width=0.496\linewidth]{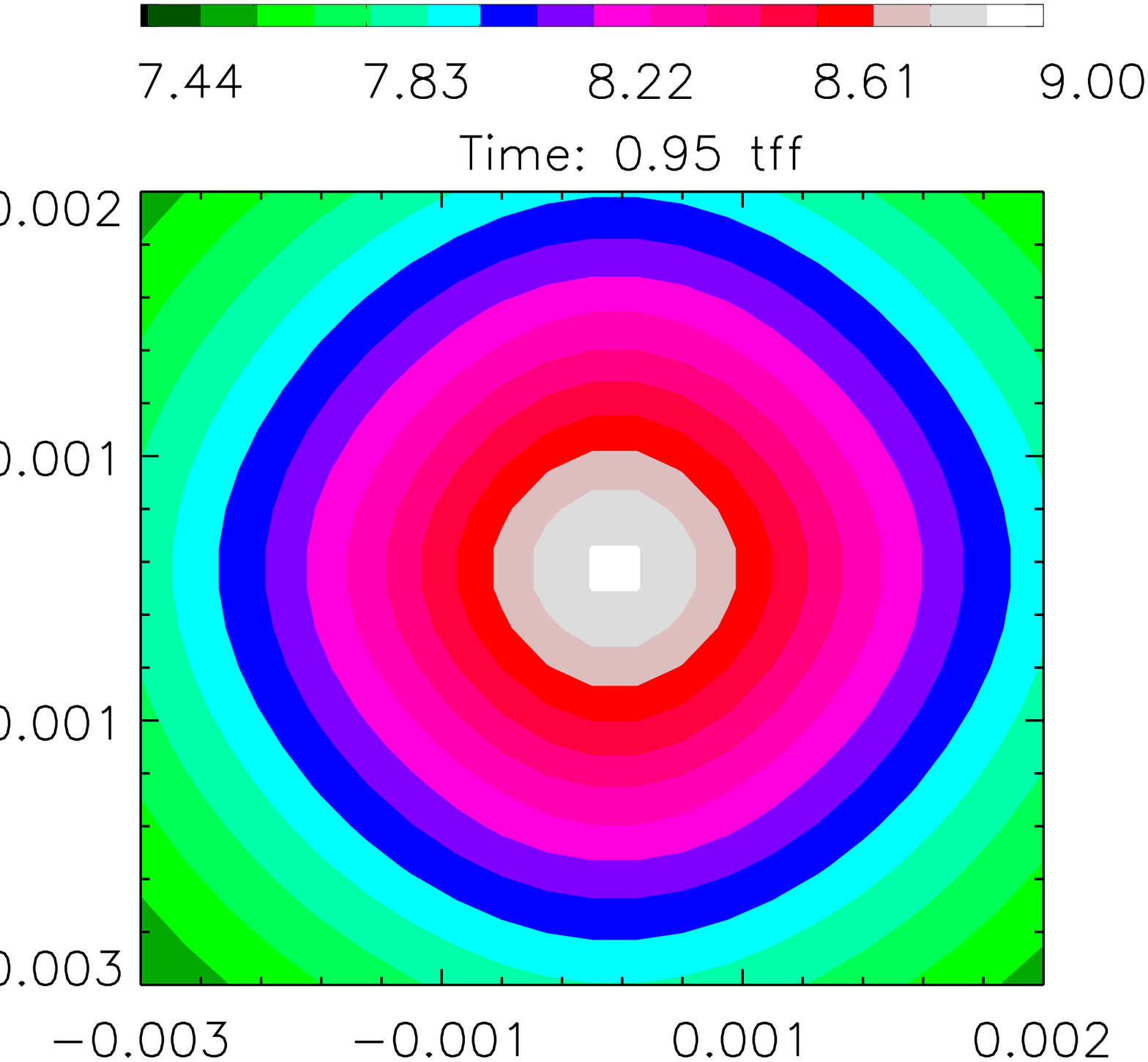}
   \includegraphics[width=0.496\linewidth]{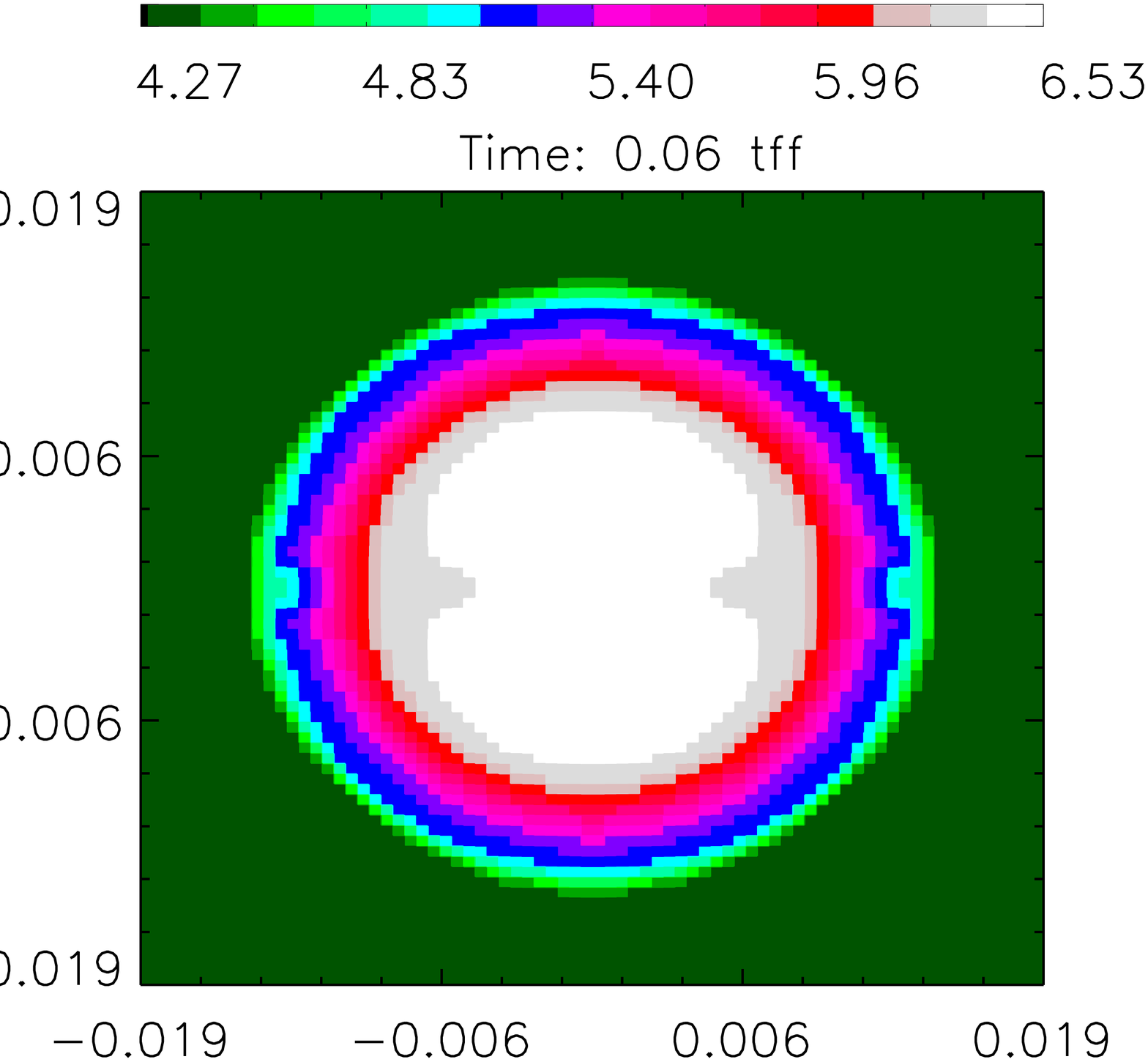}
    \includegraphics[width=0.496\linewidth]{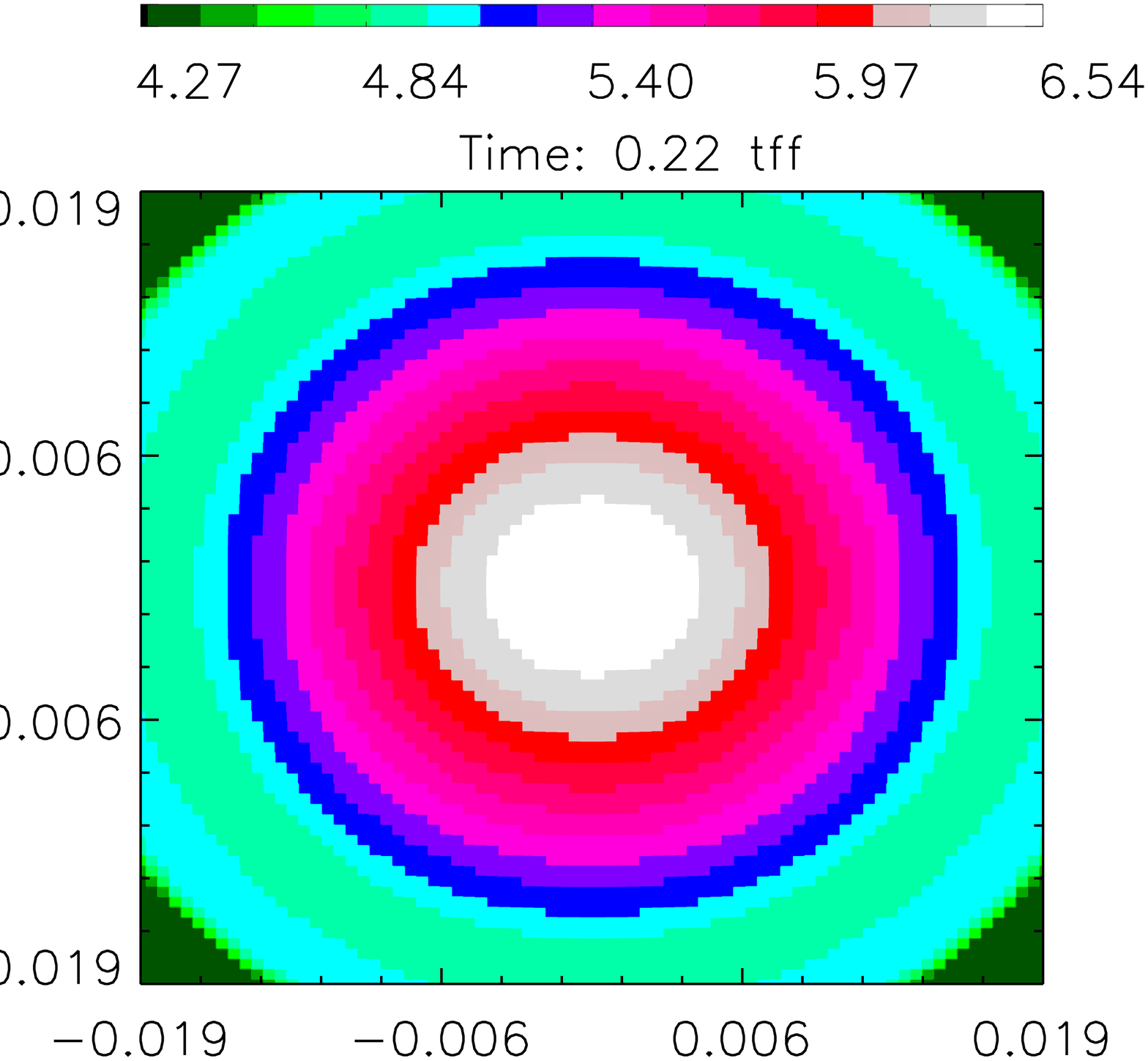}
    \caption{Contours of the log $n_H$ in units of $m_H$ cm$^{-3}$ along the yz plane. From top to bottom: $l=2$ mode for virial parameters $\alpha=0.1$, $\alpha=0.2$ and $\alpha=0.8$.  Like in Fig. \ref{a0.006}, units along each axis are parsecs and $t_{ff}$ stands for free-fall time.}
     \label{l2_num}
\end{figure}

Some examples of the core behavior are shown in Figs.~\ref{a0.006} and \ref{l2_num}, where we show density contour plots of a cut along the yz plane in the middle of the simulation box for different times of each simulation.  
For low virial parameters (largest cores, $\alpha < 0.15$) the density peaks are more and more enhanced with respect to the background,
which, as the collapse continues, can lead to the formation of as many objects as the peaks of the initial perturbation
In the intermediate virial parameter regime ( $0.15<\alpha<0.6$), the perturbations are erased by the expanding rarefaction wave and the core
collapses into one single fragment.  In the large virial parameter regime ( $0.6<\alpha$, smallest cores) the core is no longer held by gravity and re-expands under the influence of its thermal pressure.

%
\subsection{Growth rates}

The growth rate of the initial perturbation is measured as the rate of change of $\delta\rho/\rho_0$, where $\delta\rho = \rho_{max}-\rho_0$.
  Here $\rho_{max}$ is defined as the maximum density in the core and $\rho_0$ as the central density of the core at each instant.
The time is normalized as $t' = 1-t/t_{ff}$, where $t$ the actual time in the simulation and $t_{ff}$ the initial free-fall time of the core,
$t_{ff}=3\pi/32G\rho_0$, where $\rho_0$ the initial density of the core.
In practice, we plot the logarithm of these quantities and calculate a least-square fit to the linear regime of the plot.
The slope of this fit is the negative of the growth rate we are after.  To make it clearer,
if the denote the growth rate with $\sigma$, like in the analytical treatment of the previous sections, then 
$\left(\delta R/ R_0\right)=\left(\delta R/ R_0\right)_0 t'^{\sigma}$ 
(see Eq. (\ref{perturb}) for the definition of the quantities, with the exception of $t'$, which here denotes the simulation time). 
By taking the logarithm of this relation one can clearly see 
that $\sigma$ becomes the slope of a linear relation between $\log \left(\delta R/ R_0\right)$ and $\log (t')$.  But since in our notation time is negative, we must also take the negative of the slope as the growth rate of the instability.

\begin{figure}
    \includegraphics[width=\linewidth]{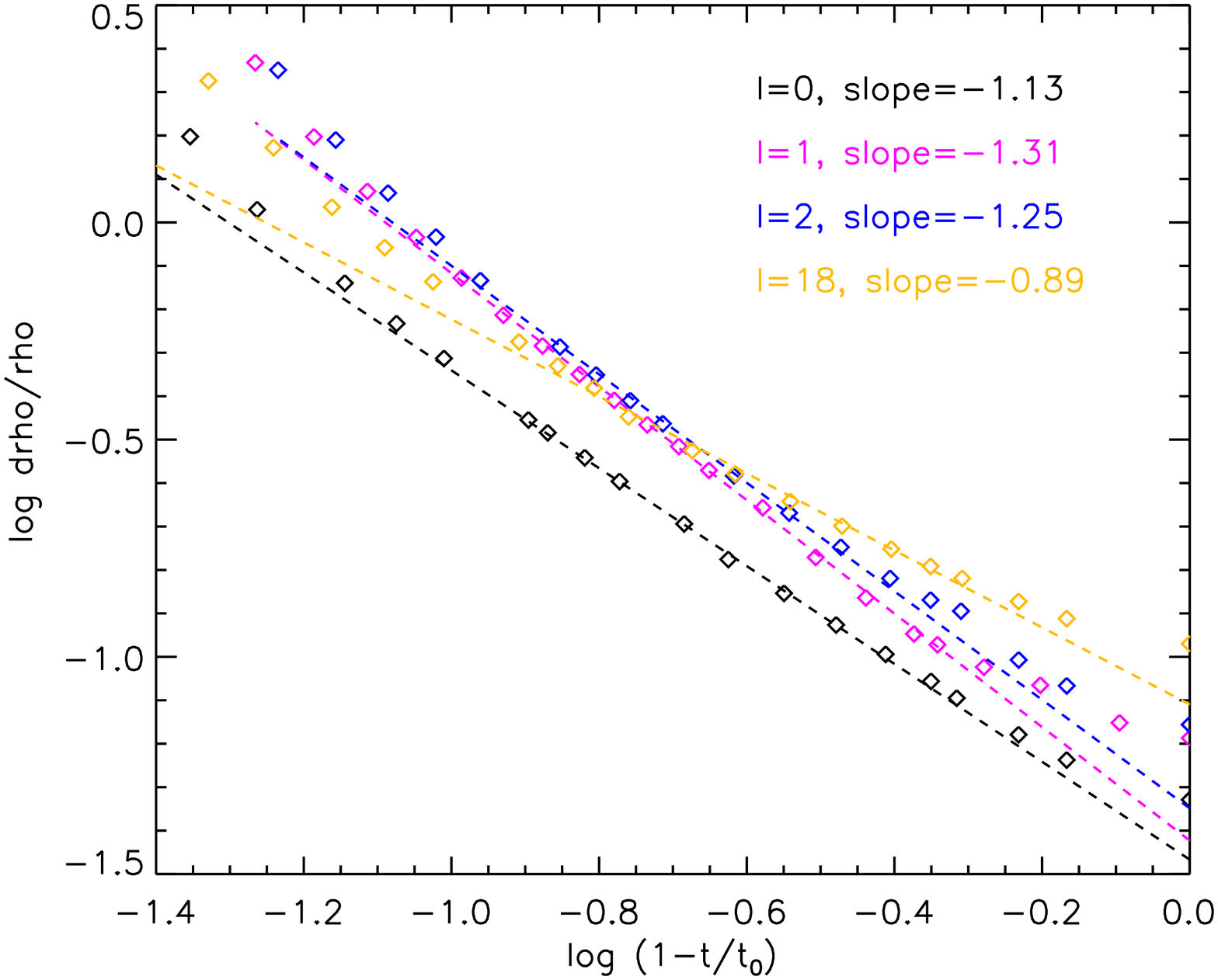}
    \caption{Growth rates for $\alpha=0.006$ mode for azimuthal wave numbers $l$.}
     \label{a0.006_multi}
\end{figure}

\begin{figure}
    \includegraphics[width=\linewidth]{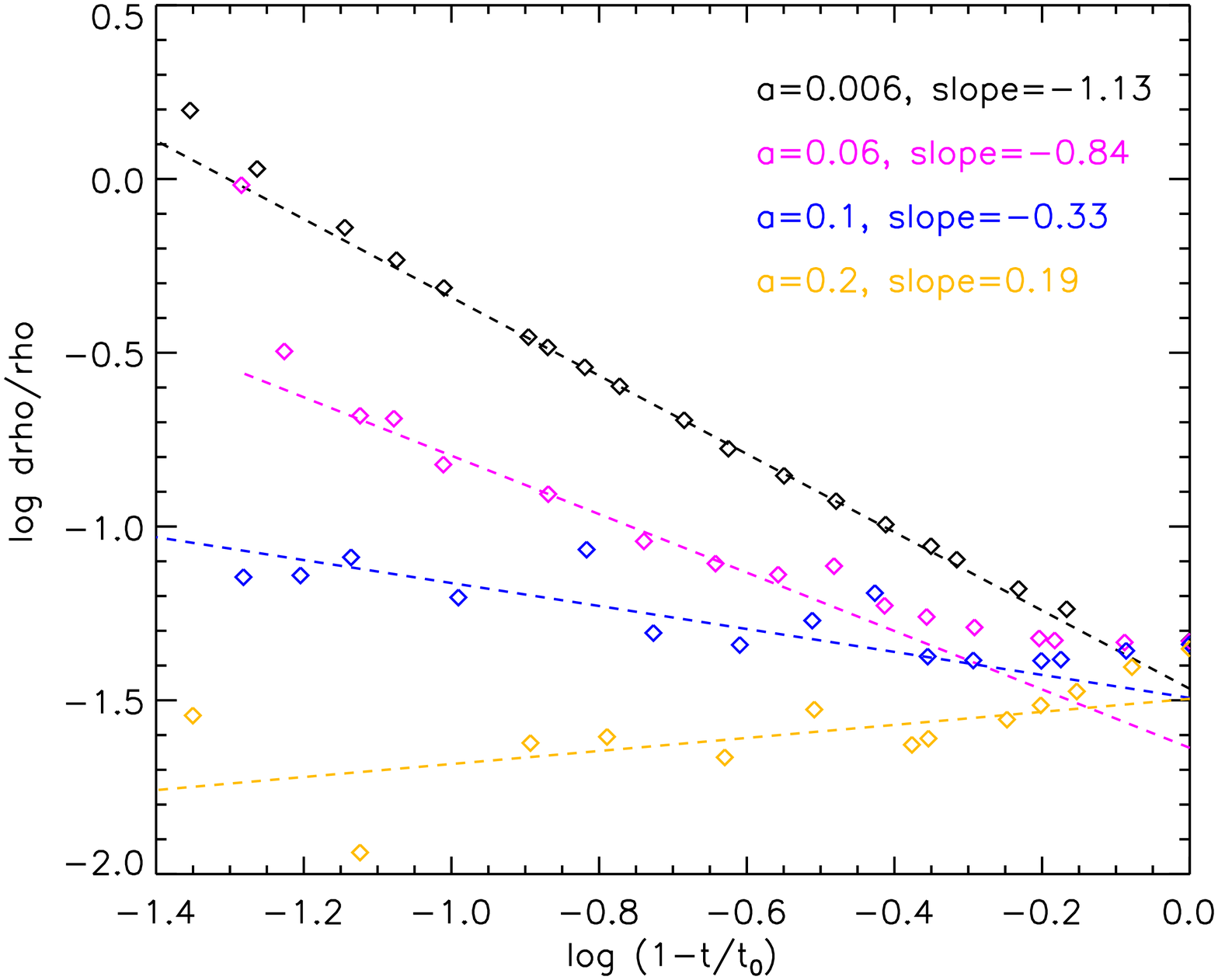}
    \caption{Growth rates of the $l=0$ mode for various virial parameters $\alpha$.}
     \label{l0_multi}
\end{figure}

\begin{figure}
    \includegraphics[width=\linewidth]{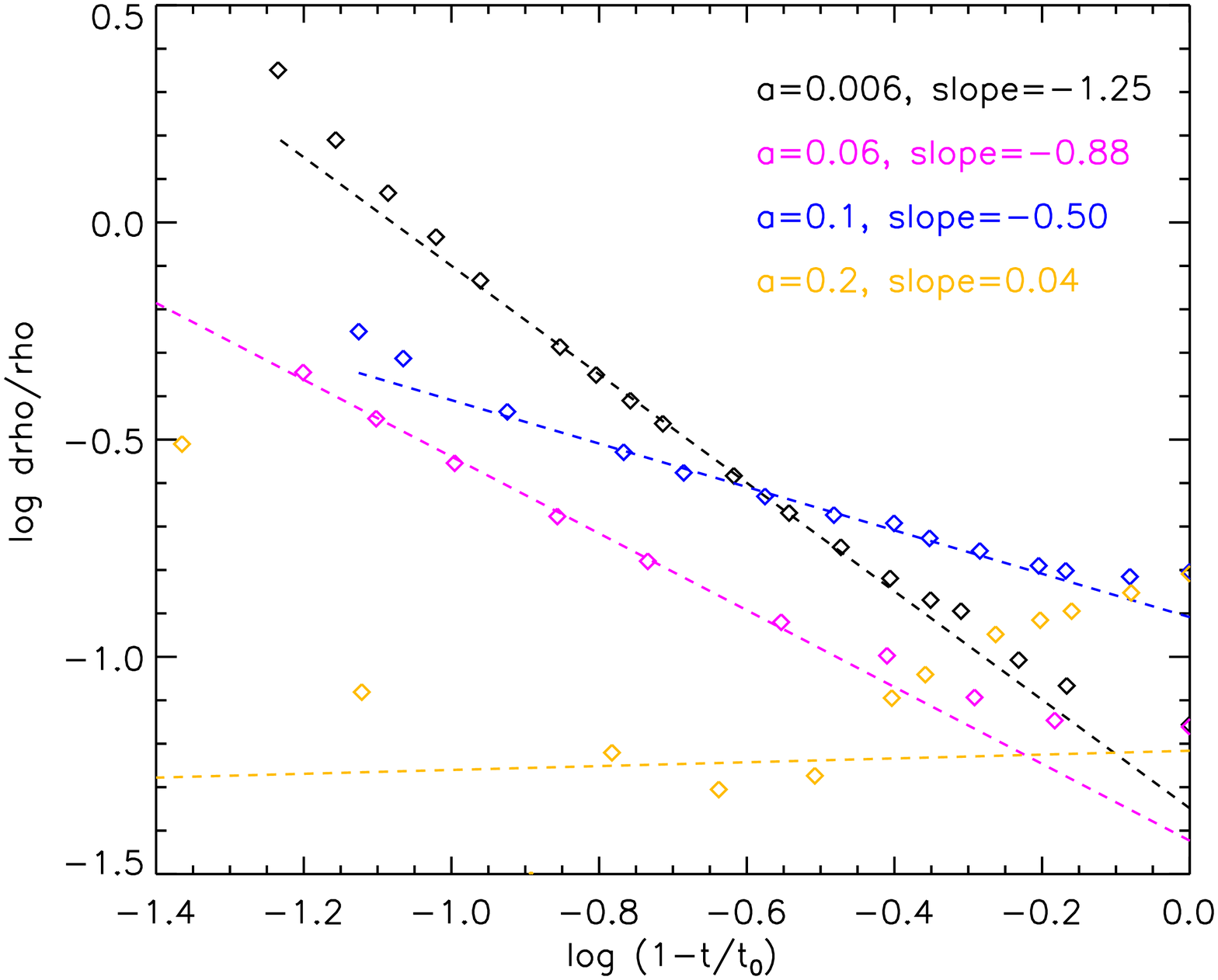}
    \caption{Growth rates of the $l=2$ mode for various virial parameters $\alpha$}
     \label{l2_multi}
\end{figure}

Some growth rates thus calculated are shown plotted in Figures \ref{a0.006_multi}, \ref{l0_multi} and \ref{l2_multi}.

Following a simple line of thought, we expect the highest growth rates to appear for the smallest $l$, since the azimuthal flow of material is then focused
on a smaller number of fragments, which can therefore grow faster. At the limit of very high $l$ on the other hand, one expects to recover $\sigma$ values similar to the $l=0$ case, since a very large number of fragments in the azimuthal direction is roughly equivalent to a continuous shell structure.

Figure ~\ref{a0.006_multi} shows the growth rates for the case of the largest core, with a virial parameter of $\alpha=0.006$. 
For the low-$l$ regime (approximately $0 < l < 10$), the growth rates inferred from the simulations are above the value of 1, in excellent agreement with the analytical prediction of the previous sections.  Also, the highest growth rates are observed for $l=1$ and $l=2$, in accordance with the simple prediction just above.  But for very high values of $l$, instead of tending towards the value of $\sigma$ for $l=0$ the growth rates in the simulations fall below unity.  This disagreement can probably be attributed to a lack of angular resolution.

Figures ~\ref{l0_multi} and ~\ref{l2_multi}, show growth rates for the same azimuthal wavenumber ($l=0$ and $l=2$, respectively), but for different virial parameters $\alpha$.  Both figures show a similar pattern: a decrease in the growth rate as the virial parameter increases, which can be understood as an increasing thermal stability of the core.  As thermal pressure becomes important, the rarefaction wave is increasingly more efficient in weakening the perturbation, to the point that, for large enough $\alpha$, it doesn't grow anymore.  

At this point we should mention that it was impossible to locate the weak shock in the simulations, due to the very limited resolution in the central parts of the core (the region with $X<3$ was only resolved with three to five cells).  This makes a direct comparison of the eigenvectors very challenging.  Nonetheless, the excellent agreement of the growth rates obtained with two such different methods is very encouraging.

\section{Discussion and implications}
\label{discussion}

The most important result of the present work is that 
a homologously collapsing cloud is prone to a shell instability due to its self-gravity, 
even for non-axisymmetric perturbations with large azimuthal wave numbers. 
This instability develops relatively fast when 
the cloud has little thermal support, showing growth rates typically 
larger than 1 and up to 1.2-1.3 for $l=2$. 
These growth rates decrease when the thermal support increases 
and eventually reach zero, meaning that the cloud is not expected 
to undergo fragmentation, at least not through the shell-like mode studied here. 

It is important to contrast this result with the analysis of HM99 for the LP solution. 
In their study, only the $l=2$ mode is unstable and 
it grows like $\simeq (t_0-t)^{-0.354}$. Since the density $\rho$ 
is proportional to $(t_0-t)^{-2}$, this implies that the perturbation grows like $\simeq \rho^{0.175}$. 
Therefore, a perturbation starting with an amplitude $\epsilon\simeq0.1$ should become
nonlinear ($\epsilon\simeq 1$), only when the density within the cloud has grown by $10^{1/0.175} \simeq 5 \times 10^5$.
This is such a significant density increase that this instability may never happen.
Take for example a dense core with a peak density of $10^5$ cm$^{-3}$. 
According to the HM99 growth rate, the density perturbation should become nonlinear only 
when the density is about $5 \times 10^{10}$ cm$^{-3}$. At these densities 
the cloud is not expected to be isothermal anymore since dust is already opaque to its radiation. 

In contrast, if the core is cold enough, the shell mode may grow as 
$\simeq (t_0-t)^{-1}$ and therefore  as $\simeq \rho^{0.5}$
This would imply that a 10$\%$ amplitude perturbation in a core of mean particle density of $10^5$ {\text{cm}}$^{-3}$  becomes 
nonlinear for a density of about $10^7$  ${\text{cm}}^{-3}$, which is entirely reasonable. 

Since the LP flow also describes a very fast collapse, the question arises
why do the two types of solutions present such different fragmentation properties? 
We believe the answer relies on the tidal forces, which differ much in the two cases {(see also \citet{Jog_2013}
for an analysis of the effect of tidal forces on the Jeans stability criterion).}

To illustrate this, let us consider a core whose density profile is $\rho=A r^{-p}$ and a
perturbation $\delta \rho$ of characteristic size $\delta r$. 
A fluid particle located at the edge of the perturbation, i.e. at a distance $\delta r$ from the perturbation center,
  is subject to the perturbed gravitational force, 
\begin{eqnarray}
\delta F_{sg} \simeq -G \delta M / \delta r ^2,
\end{eqnarray}
where $\delta M = 4 \pi/3 \delta r^3 \rho$, 
as well as to the tidal forces, that is the gradient of the gravitational force produced by the 
mean density. The gravitational acceleration within the core is $a_g = GM(r) / r^2$
with $M=(4 \pi/3) (A / (3-p)) r^{3-p}$ and 
\begin{eqnarray}
a_g = - {4 \pi \over 3} G { A \over  3-p} r^{1-p}.
\end{eqnarray}
Thus the tidal force is 
\begin{eqnarray}
\delta F_{t}=\delta \rho (a_g(r+ \delta r)-a_g(r))=- {4 \pi \over 3 } \rho \delta \rho G { 1-p \over 3-p } \delta r.
\end{eqnarray}
The fluid particle at the edge of the perturbation feels a total gravitational force equal to
\begin{eqnarray}
\delta F_{tot} =\delta  F_{sg} +\delta F_{t} =  {4- 2p \over  3-p}  \delta F_{sg}.
\end{eqnarray}
Thus the tidal force modifies the effective gravity of the perturbation.  We can identify the following regimes for the values of the exponent 
$p$ in the equilibrium profile:
\begin{itemize}
\item If $p=0$, then $\delta F_{tot}  =  4/3  d F_{sg}$, which means that the tidal force enhances the 
gravitational instability because it compresses the perturbation. 

\item If $p=1$, then $\delta F_{tot}  =  d F_{sg}$ and the tidal force has no effect. 

\item If $p=1.5$, then $\delta F_{tot}  =  2/3 d F_{sg}$, in other words the tidal force works against total gravity because 
it tends to shear the fluid elements apart. 

\item Finally, if $p=2$, then $\delta F_{tot}  =  0$ and the tidal force cancels the gravitational force of
the perturbation. In this case, only nonlinear or non-axisymetric perturbations can develop. 
\end{itemize}

This short analysis suggests that the homologously collapsing, uniform density solution is more prone to the gravitational 
shell instability that the Larson-Penston solution, which presents an $r^{-2}$ density profile at large radii.  This result is also in excellent
agreement with both the analytical and the numerical findings presented in this work.

Our results concerning the role of the equilibrium profile in the fate of the perturbations 
are also compatible with the numerical simulations performed by \citet{Girichidis_11}.  
These authors performed a series of simulations of massive turbulent cores, in which they vary the initial density profiles
of the clouds. In particular, they found that when the density is initially uniform or given by a Bonnor-Ebert density 
profile, the cloud fragments in many objects. On the contrary, 
if the density profile is initially proportional to $r^{-2}$, only one or very few fragments form.  
This is entirely consistent with the above interpretation of the tidal forces. 

In terms of understanding the origin of stellar multiplicity, where do the above results leave us?
In short, our study complements a large volume of previous literature on the subject of core fragmentation, 
which so far has been mostly dealing with rotating or turbulent environments.  This, of course, is justified
by the very structured kinematics typically observed in pre-stellar cores \citep[for a partial review]{Goodwin_2007}, which 
very often do show irrefutable rotation signatures \citep{Goodman_1993}. 

And although there is little doubt that rotation does lead to fragmentation, 
here we suggest a "thermal" type of fragmentation that does not require any initial rotation.
In a large and cold enough cloud, if its density profile is flat, perturbations will grow to become nonlinear, eventually leading to fragmentation.
The fragments, unlike in cases with rotation, which produce tens to hundreds of AU separations, \citep[for example]{Hennebelle_2004, Commercon_2008} will be located at distances of more than 1000~AU during the first core collapse, with a possibility of migrating inward at later stages and, with the onset of rotation, become binaries, like observed, for example in the simulations of \citet{Hennebelle_Teyssier_08} for magnetized rotating environments.

Although limited in their application due to the idealized character of the setup, the results of this study are useful, both for
predicting the behavior of initially flat, quiescent cores at a stage of their collapse almost inaccessible observationally, but also, very importantly, fot deciphering the behavior of more complex, dynamical models.

\section{Conclusions}

We have calculated the form of linear perturbations for the self-similar flow that describes the homologous collapse of an isothermal sphere. 
The calculation was done by introducing a weak shock at the sonic point of the flow and integrating the corresponding perturbation equations from   location to infinity with a Runge-Kutta scheme.  The problem was treated in one dimension for the case of a spherical shell and in three dimensions for non-axisymmetric perturbations.  Results of this analysis follow.
\begin{itemize}
\item Using a shock at the critical point of the self-similar profile allows the calculation of physically meaningful perturbations.
\item In the spherical case, physically acceptable solutions form a continuum with growth rates $\sigma>1$.
\item The non-spherical modes exhibit growth rates similar to the shell mode.  This implies that it is the shell mode that 
drives the instability, with the higher order modes growing on top of it and providing the multiplicity of the fragments.
\end{itemize}

In parallel, we performed direct numerical simulations of a collapsing uniform isothermal sphere with a linear perturbation, varying the degree of thermal support (quantified by the ratio of thermal to gravitational energy, or virial parameter $\alpha$) and the shape of the initial perturbation.
For the case of smallest thermal support we tried ($\alpha=0.006$), the growth rates estimated from the simulations are in perfect accordance to the analytical expectations.

By increasing the degree of thermal support, we find that the density peaks grow slower and slower with respect to the background, while for larger values of the virial parameter the density contrast decreases with time.  This suggests that eventually, non-rotating cores with $0.15 < \alpha < 0.6$ would collapse to 
a single object, while for larger values of $\alpha$ they would re-expand.

Our results suggest the existence of a "thermal" type of fragmentation for cold, large clouds 
that does not require the presence of initial rotation and that can produce very widely separated fragments at
the early stages of the collapse.

\emph{Acknowledgments}
We are grateful to Philippe Andr\'{e}, Anaelle Maury, Tomoyuki Hanawa and Shu-ichiro Inutsuka for their comments and suggestions.  
We also thank the anonymous referee for useful comments that helped improve this manuscript.
This work is largely a result of our participation in the International Summer Institute for
Modeling in Astrophysics (ISIMA) during the summer of 2011.  
We especially thank Pascale Garaud for organizing the meeting and for useful discussions.
This research has received funding from the European Research Council under the European Community's Seventh Framework Programme 
(ERC Grant Agreement "ORISTARS", no. 291294 and FP7/2007-2013 Grant Agreement no. 306483). 

\bibliographystyle{apj}
\bibliography{collap_stab}

\begin{thebibliography}{50}
\expandafter\ifx\csname natexlab\endcsname\relax\def\natexlab#1{#1}\fi

\bibitem[{{Andre} {et~al.}(1993){Andre}, {Ward-Thompson}, \&
  {Barsony}}]{Andre_93}
{Andre}, P., {Ward-Thompson}, D., \& {Barsony}, M. 1993, \apj, 406, 122

\bibitem[{{Andre} {et~al.}(2000){Andre}, {Ward-Thompson}, \&
  {Barsony}}]{Andre_00}
---. 2000, Protostars and Planets IV, 59

\bibitem[{{Banerjee} \& {Pudritz}(2006)}]{Banerjee_Pudritz_2006}
{Banerjee}, R., \& {Pudritz}, R.~E. 2006, \apj, 641, 949

\bibitem[{{Bate}(2009)}]{Bate_09}
{Bate}, M.~R. 2009, \mnras, 397, 232

\bibitem[{{Bonnor}(1956)}]{Bonnor_1956}
{Bonnor}, W.~B. 1956, \mnras, 116, 351

\bibitem[{{Boss} \& {Bodenheimer}(1979)}]{Boss_Bodenheimer_1979}
{Boss}, A.~P., \& {Bodenheimer}, P. 1979, \apj, 234, 289

\bibitem[{{Boss} \& {Keiser}(2013)}]{Boss_Keiser_2013}
{Boss}, A.~P., \& {Keiser}, S.~A. 2013, \apj, 764, 136

\bibitem[{{Burkert} \& {Bodenheimer}(1993)}]{Bodenheimer_Burkert_1993}
{Burkert}, A., \& {Bodenheimer}, P. 1993, \mnras, 264, 798

\bibitem[{{Cha} \& {Whitworth}(2003)}]{Cha_Whitworth_2003}
{Cha}, S.-H., \& {Whitworth}, A.~P. 2003, \mnras, 340, 91

\bibitem[{{Chen} {et~al.}(2008){Chen}, {Launhardt}, {Bourke}, {Henning}, \&
  {Barnes}}]{Chen_08}
{Chen}, X., {Launhardt}, R., {Bourke}, T.~L., {Henning}, T., \& {Barnes}, P.~J.
  2008, \apj, 683, 862

\bibitem[{{Commer{\c c}on} {et~al.}(2008){Commer{\c c}on}, {Hennebelle},
  {Audit}, {Chabrier}, \& {Teyssier}}]{Commercon_2008}
{Commer{\c c}on}, B., {Hennebelle}, P., {Audit}, E., {Chabrier}, G., \&
  {Teyssier}, R. 2008, \aap, 482, 371

\bibitem[{{Duch{\^e}ne} {et~al.}(2007){Duch{\^e}ne}, {Bontemps}, {Bouvier},
  {Andr{\'e}}, {Djupvik}, \& {Ghez}}]{Duchene_2007}
{Duch{\^e}ne}, G., {Bontemps}, S., {Bouvier}, J., {et~al.} 2007, \aap, 476, 229

\bibitem[{{Duch{\^e}ne} {et~al.}(2004){Duch{\^e}ne}, {Bouvier}, {Bontemps},
  {Andr{\'e}}, \& {Motte}}]{Duchene_2004}
{Duch{\^e}ne}, G., {Bouvier}, J., {Bontemps}, S., {Andr{\'e}}, P., \& {Motte},
  F. 2004, \aap, 427, 651

\bibitem[{{Duquennoy} \& {Mayor}(1991)}]{Duquennoy_Mayor_1991}
{Duquennoy}, A., \& {Mayor}, M. 1991, \aap, 248, 485

\bibitem[{{Ebert}(1955)}]{Ebert_1955}
{Ebert}, R. 1955, \zap, 37, 217

\bibitem[{{Falgarone} \& {Phillips}(1990)}]{Falgarone_Phillips_1990}
{Falgarone}, E., \& {Phillips}, T.~G. 1990, \apj, 359, 344

\bibitem[{{Ghez} {et~al.}(1993){Ghez}, {Neugebauer}, \& {Matthews}}]{Ghez_93}
{Ghez}, A.~M., {Neugebauer}, G., \& {Matthews}, K. 1993, \aj, 106, 2005

\bibitem[{{Girart} {et~al.}(2009){Girart}, {Rao}, \& {Estalella}}]{Girart_09}
{Girart}, J.~M., {Rao}, R., \& {Estalella}, R. 2009, \apj, 694, 56

\bibitem[{{Girichidis} {et~al.}(2011){Girichidis}, {Federrath}, {Banerjee}, \&
  {Klessen}}]{Girichidis_11}
{Girichidis}, P., {Federrath}, C., {Banerjee}, R., \& {Klessen}, R.~S. 2011,
  \mnras, 413, 2741

\bibitem[{{Goodman} {et~al.}(1993){Goodman}, {Benson}, {Fuller}, \&
  {Myers}}]{Goodman_1993}
{Goodman}, A.~A., {Benson}, P.~J., {Fuller}, G.~A., \& {Myers}, P.~C. 1993,
  \apj, 406, 528

\bibitem[{{Goodwin} {et~al.}(2007){Goodwin}, {Kroupa}, {Goodman}, \&
  {Burkert}}]{Goodwin_2007}
{Goodwin}, S.~P., {Kroupa}, P., {Goodman}, A., \& {Burkert}, A. 2007,
  Protostars and Planets V, 133

\bibitem[{{Hachisu} \& {Eriguchi}(1984)}]{Hachisu_Eriguchi_1984}
{Hachisu}, I., \& {Eriguchi}, Y. 1984, \aap, 140, 259

\bibitem[{{Hanawa} \& {Matsumoto}(1999)}]{Hanawa_Matsumoto_1999}
{Hanawa}, T., \& {Matsumoto}, T. 1999, \apj, 521, 703

\bibitem[{{Hennebelle} \& {Teyssier}(2008)}]{Hennebelle_Teyssier_08}
{Hennebelle}, P., \& {Teyssier}, R. 2008, \aap, 477, 25

\bibitem[{{Hennebelle} {et~al.}(2004){Hennebelle}, {Whitworth}, {Cha}, \&
  {Goodwin}}]{Hennebelle_2004}
{Hennebelle}, P., {Whitworth}, A.~P., {Cha}, S.-H., \& {Goodwin}, S.~P. 2004,
  \mnras, 348, 687

\bibitem[{{Hunter}(1964)}]{Hunter_1964}
{Hunter}, C. 1964, \apj, 139, 570

\bibitem[{{Jog}(2013)}]{Jog_2013}
{Jog}, C.~J. 2013, \mnras, 434, L56

\bibitem[{{Joos} {et~al.}(2013){Joos}, {Hennebelle}, {Ciardi}, \&
  {Fromang}}]{Joos_13}
{Joos}, M., {Hennebelle}, P., {Ciardi}, A., \& {Fromang}, S. 2013, \aap, 554,
  A17

\bibitem[{{J{\o}rgensen} {et~al.}(2007){J{\o}rgensen}, {Bourke}, {Myers}, {Di
  Francesco}, {van Dishoeck}, {Lee}, {Ohashi}, {Sch{\"o}ier}, {Takakuwa},
  {Wilner}, \& {Zhang}}]{Jorgensen_07}
{J{\o}rgensen}, J.~K., {Bourke}, T.~L., {Myers}, P.~C., {et~al.} 2007, \apj,
  659, 479

\bibitem[{{Klessen} \& {Burkert}(2001)}]{Klessen_01}
{Klessen}, R.~S., \& {Burkert}, A. 2001, \apj, 549, 386

\bibitem[{{Klessen} {et~al.}(1998){Klessen}, {Burkert}, \&
  {Bate}}]{Klessen_1998}
{Klessen}, R.~S., {Burkert}, A., \& {Bate}, M.~R. 1998, \apjl, 501, L205

\bibitem[{{Larson}(1969)}]{Larson_1969}
{Larson}, R.~B. 1969, \mnras, 145, 271

\bibitem[{{Larson}(1981)}]{Larson_1981}
---. 1981, \mnras, 194, 809

\bibitem[{{Machida} {et~al.}(2008){Machida}, {Omukai}, {Matsumoto}, \&
  {Inutsuka}}]{Machida_2008}
{Machida}, M.~N., {Omukai}, K., {Matsumoto}, T., \& {Inutsuka}, S.-i. 2008,
  \apj, 677, 813

\bibitem[{{Matsumoto} \& {Hanawa}(2003)}]{Matsumoto_Hanawa_03}
{Matsumoto}, T., \& {Hanawa}, T. 2003, \apj, 595, 913

\bibitem[{{Maury} {et~al.}(2010){Maury}, {Andr{\'e}}, {Hennebelle}, {Motte},
  {Stamatellos}, {Bate}, {Belloche}, {Duch{\^e}ne}, \& {Whitworth}}]{Maury_10}
{Maury}, A.~J., {Andr{\'e}}, P., {Hennebelle}, P., {et~al.} 2010, \aap, 512,
  A40

\bibitem[{{Miyama} {et~al.}(1984){Miyama}, {Hayashi}, \&
  {Narita}}]{Miyama_1984}
{Miyama}, S.~M., {Hayashi}, C., \& {Narita}, S. 1984, \apj, 279, 621

\bibitem[{{Myhill} \& {Kaula}(1992)}]{Myhill_Kaula_1992}
{Myhill}, E.~A., \& {Kaula}, W.~M. 1992, \apj, 386, 578

\bibitem[{{Offner} {et~al.}(2008){Offner}, {Klein}, \& {McKee}}]{Offner_08}
{Offner}, S.~S.~R., {Klein}, R.~I., \& {McKee}, C.~F. 2008, \apj, 686, 1174

\bibitem[{{Ori} \& {Piran}(1988)}]{Ori_Piran_1988}
{Ori}, A., \& {Piran}, T. 1988, \mnras, 234, 821

\bibitem[{{Patience} {et~al.}(2002){Patience}, {White}, {Ghez}, {McCabe},
  {McLean}, {Larkin}, {Prato}, {Kim}, {Lloyd}, {Liu}, {Graham}, {Macintosh},
  {Gavel}, {Max}, {Bauman}, {Olivier}, {Wizinowich}, \&
  {Acton}}]{Patience_2002}
{Patience}, J., {White}, R.~J., {Ghez}, A.~M., {et~al.} 2002, \apj, 581, 654

\bibitem[{{Penston}(1969)}]{Penston_1969}
{Penston}, M.~V. 1969, \mnras, 144, 425

\bibitem[{{Price} \& {Bate}(2007)}]{Price_Bate_2007}
{Price}, D.~J., \& {Bate}, M.~R. 2007, \mnras, 377, 77

\bibitem[{{Shu}(1977)}]{Shu_1977}
{Shu}, F.~H. 1977, \apj, 214, 488

\bibitem[{{Teyssier}(2002)}]{Teyssier_02}
{Teyssier}, R. 2002, \aap, 385, 337

\bibitem[{{Tohline}(1981)}]{Tohline_1981}
{Tohline}, J.~E. 1981, \apj, 248, 717

\bibitem[{{Tohline}(2002)}]{Tohline_2002}
---. 2002, \araa, 40, 349

\bibitem[{{Truelove} {et~al.}(1997){Truelove}, {Klein}, {McKee}, {Holliman},
  {Howell}, \& {Greenough}}]{Truelove_1997}
{Truelove}, J.~K., {Klein}, R.~I., {McKee}, C.~F., {et~al.} 1997, \apjl, 489,
  L179

\bibitem[{{Tsuribe} \& {Inutsuka}(1999)}]{Tsuribe_Inutsuka_1999}
{Tsuribe}, T., \& {Inutsuka}, S.-I. 1999, \apj, 526, 307

\bibitem[{{Whitworth} \& {Summers}(1985)}]{Whitworth_Summers_1985}
{Whitworth}, A., \& {Summers}, D. 1985, \mnras, 214, 1

\end{thebibliography}
\label{lastpage}

\end{document}